# A comprehensive study of comet 67P/Churyumov-Gerasimenko in the 2021/2022 apparition. I. Photometry, spectroscopy, morphology

Vira Rosenbush [a,b], Valerii Kleshchonok [a,c], Oleksandra Ivanova [a,b,d], Igor Luk'yanyk [a,*],
Colin Snodgrass [e], Daniel Gardener [e], Ludmilla Kolokolova [f], Johannes Markkanen [g],
Elena Shablovinskaya [h]

[a] *Astronomical Observatory of Taras Shevchenko National University of Kyiv, 3 Observatorna St., Kyiv 04053, Ukraine*
[b] *Main Astronomical Observatory of the National Academy of Sciences of Ukraine, 27 Akademika Zabolotnoho St., Kyiv 03143, Ukraine*
[c] *Max Planck Institute for Solar System Research, Justus-von-Liebig-Weg 3, D-37077 Göttingen, Germany*
[d] *Astronomical Institute of Slovak Academy of Sciences, 059 60 Tatranska Lomnica, Slovak Republic*
[e] *Institute for Astronomy, University of Edinburgh, Edinburgh EH9 3HJ, UK*
[f] *University of Maryland, College Park, MD, USA*
[g] *Institut für Geophysik und Extraterrestrische Physik, Technische Universität Braunschweig, Mendelssohnstr. 3, D-38106 Braunschweig, Germany*
[h] *Núcleo de Astronomá de la Facultad de Ingenierá, Universidad Diego Portales, Av. Ejército Libertador 441, Santiago, Chile*

## ABSTRACT

We present observations of comet 67P/Churyumov–Gerasimenko during its 2021/22 apparition, aiming to investigate its dust and gas environment and compare the results with those obtained in 2015/16 using the same telescope. Quasi-simultaneous photometric, spectroscopic, and polarimetric observations were carried out at the 6-m BTA SAO telescope. The comet was observed on 6 October 2021, 31 days before perihelion, with g-sdss and r-sdss filters, and on 6 February 2022, 96 days after perihelion, using narrowband cometary filters: $BC$ ($\lambda 4450/62$ Å), $RC$ ($\lambda 6839/96$ Å), and $CN$ ($\lambda 3870/58$ Å). These were complemented by images from the 2-m Liverpool Telescope (La Palma). On 6 October 2021, a sunward jet and long dust tail were detected. By 6 February 2022, the dust coma morphology had changed noticeably, revealing a bright sunward neckline structure superimposed on the projected dust tail, along with two jets at position angles of $133°$ and $193°$. Spectra showed strong CN emission, with relatively weak $C_2$, $C_3$, and $NH_2$ emissions. The dust production rate $Af\rho$ did not exceed 200 cm (uncorrected for phase angle) in both epochs. An unusual CN coma morphology was observed, with evidence of an additional CN source associated with dust jets. Geometric modeling of the jets' dynamics indicated an active area at latitude $-70° \pm 4°$ with a jet opening angle of $20° \pm 6°$ on 6 October 2021, and two active areas at latitudes $-58° \pm 5°$ and $-53° \pm 10°$, separated by longitude $150° \pm 20°$, producing the observed jets on 6 February 2022. The average particle velocity in the jets was about $0.32 \pm 0.04$ km s$^{-1}$.

*Keywords:*
Comets: general; Comets: individual; 67P/Churyumov–Gerasimenko Methods: miscellaneous Techniques: imaging spectroscopy Techniques: photometric

## 1. Introduction

Comet 67P/Churyumov– Gerasimenko (67P/C– G hereafter) is a Jupiter family comet with an orbital period of about 6.44 years. In the previous apparition, the comet was intensively studied by the European Space Agency's *Rosetta* rendezvous mission from 2014 to 2016, as well as a worldwide ground-based campaign; thus, it is one of the best- studied comets. In that apparition, within the framework of the ground-based support of the *Rosetta* mission, we carried out extensive observations of comet 67P/C– G in the post-perihelion period from November 2015 to April 2016 at a heliocentric distance from 1.61 au to 2.72 au using the 6-m Big Telescope Alt-azimuth (BTA) telescope of the Special Astrophysical Observatory (SAO). These observations provided valuable insights into coma processes beyond Rosetta's coverage (Ivanova et al., 2017a; Rosenbush et al., 2017; Snodgrass et al., 2017; Boehnhardt et al., 2024).

Comet 67P/C– G passed its last perihelion on 2.1 November 2021 UT at a heliocentric distance of $r = 1.211$ au. In this apparition, the comet was at the closest approach to Earth since 1982: the minimum geocen-tric distance was $\Delta = 0.418$ au on 11 November 2021. We carried out two sets of quasi-simultaneous photometric, spectroscopic, and polarimetric observations before and after perihelion using the 6-m BTA telescope. In addition, we supplemented these observations with images of the comet obtained at the robotic 2-m Liverpool Telescope (LT) on La Palma. This unique dataset can provide important information about the composition, structure, and evolutionary processes in comet 67P/C− G.

So far, there are few publications on observations of comet 67P/C− G in the 2021/22 apparition. Kelley et al. (2021) reported an apparent outburst of the comet on 17.86 November 2021 UT. This outburst was confirmed in subsequent observations on November 18. Sharma et al. (2021) later clarified that there were two outbursts: 29.94 October and 17.86 November 2021. Sharma et al. concluded that these outbursts were an order of magnitude larger than the strongest outburst observed in situ by the *Rosetta* spacecraft in 2015. Gardener et al. (2022) examined the activity levels of the comet at different parts of its orbit in the 2015/16 and 2021/22 apparitions and confirmed the conclusion by Sharma et al. (2021). In addition, the comet was active at a similar orbit point around perihelion in the 2015/16 apparition. Observations by Bair et al. (2022) showed a large pre− /post-perihelion asymmetry in production rates of gas molecules in comet 67P/C− G, ranging from about 2 to 6 times after perihelion than before, and, surprisingly, dust production rates 1.5–2.5 times before and around perihelion during the 2021/22 apparition as compared to previous ones.

In this paper, we focused on studying the optical and physical properties of the gas and dust components, as well as the morphology of the coma of comet 67P/C− G in the 2021/22 apparition using photometric and spectral data. The results of the colorimetric and polarimetric observations of the comet and their modeling will be presented in the next paper. The paper is organized as follows: Section 2 describes the instruments, observations, and data reduction processes. Optical spectra of the comet and molecular composition are presented in Section 3. The morphology of the coma together with numerical simulation of the structure of the coma, is provided in Section 4. Dust characteristics acquired from the photometric observations of the comet are described in Section 5. In Section 6, we compare the results obtained in the 2015/16 and 2021/22 apparitions, followed by the conclusions in Section 7.

## 2. Observations and reduction

In the recent 2021/22 apparition of comet 67P/C−G, we used two telescopes with different equipment for observations. Spectra and imaging photometry were acquired at the 6-m (*f/*4) BTA telescope. Also, the robotic 2-m Liverpool Telescope (LT) on La Palma was used for photometric observations of the comet, which were described by Gardener et al. (2022). In all cases, guiding the telescope at the apparent velocity of the comet was applied to the observations to compensate for proper motion during the exposures. The following briefly describes the observations and the reduction steps applied to all the science images.

### 2.1. Observations at the 6-m BTA

Quasi-simultaneous photometric and spectroscopic observations of comet 67P/C−G were carried out on 6 October 2021 and 6 February 2022 (see Table 1). Observations were conducted 31 days before and 96 days after its perihelion passage (2 November 2021). The multi-mode focal reducer SCORPIO-2 (Spectral Camera with Optical Reducer for Photometrical and Interferometrical Observations, see Afanasiev and Moiseev, 2011; Afanasiev and Amirkhanyan, 2012), mounted in the primary focus of the 6-m BTA telescope, was used in the photometric and spectroscopic modes. The seeing (FWHM) was $\sim2''$ (689 km at the distance of the comet) on 6 October 2021 and $\sim3''$ (1028 km) on 6 February 2022, respectively. All the observations were performed with a CCD detector E2V CCD261–84 of $4096 \times 2048$ pixels and a $15 \times 15$-µm pixel size. The full field of view of the CCD chip is $6.1' \times 6.1'$ with an image scale of 0.4 arcsec px$^{-1}$. The twilight sky was observed to define flat-field corrections of the photometric and polarimetric images. To increase the measured signal-to-noise (S/N) ratio, on-chip binning $2 \times 2$ was *applied* to all observed *images* in the photometric mode, and binning $1 \times 2$ in the long-slit mode. As a result, direct images (hereinafter referred to as Ima) and long-slit spectra (Sp) were obtained (see Table 1). The pre-processing of the raw images was performed with the specialized software packages (Afanasiev and Amirkhanyan, 2012). The sky-subtracted and flat-fielded individual frames were aligned with each other and summed with the robust averaging algorithm. A composite image in each SCORPIO-2 mode was acquired for each night. More detailed information is given by Afanasiev et al. (2017), Ivanova et al. (2017b), and Rosenbush et al. (2020).

#### 2.1.1. Photometry

Photometric observations of the comet were performed through the broadband filters *g*-sdss ($\lambda$4650/1300 Å) and *r*-sdss ($\lambda$6200/1200 Å) of the SDSS (Sloan Digital Sky Survey) photometric system on 6 October 2021 and the narrowband comet filters designed to isolate gas emission band *CN* ($\lambda$3870/58 Å) and two points of the continuum, blue *BC* ($\lambda$4450/62 Å) and red *RC* ($\lambda$6839/96 Å), on 6 February 2022. The sky background level was estimated in each image using regions outside the coma and tail free of faint stars. The photometric measurements made with the broadband filters in October 2021 were calibrated using the field stars as standards for each observation set. The stellar magnitudes of the standard stars were taken from the catalog AAVSO Photometric All Sky Survey (APASS[1]), which is a *compiled all-sky star catalog* covering mainly the 10 to 17 magnitude range in a single bandpass, *B, V, g', r', i'*. The photometric uncertainty of the catalog depends on star brightness and is estimated to be 0.02 mag on average. Unfortunately, in February 2022, the photometric standard was not observed with narrowband comet filters (*CN, BC,* and *RC*). Therefore, to obtain absolute fluxes, we used spectral observations made on the same night as the photometric ones, taking into account the difference in fluxes between the rectangular slit of the spectrograph and the circular aperture. The primary reductions of the images obtained were performed using a standard image processing technique, including dividing the science frames by flat fields after bias subtraction and removing cosmic rays.

#### 2.1.2. Spectroscopy

We used the VPHG1200@540 grism and the long slit with dimensions of $6.8' \times 1.0''$ for spectral observations of comet 67P/C−G during the night on 6 October 2021 and 6 February 2022. The spectra cover the effective wavelength region of $\lambda$3600–7400 Å with a mean reciprocal dispersion of 0.91 Å px$^{-1}$. The spectral resolution across the full range of wavelengths is about 5.2 Å, and the spatial scale is 0.4 arcsec px$^{-1}$ along the slit. The spectrograph slit was centered on the optocenter (see Figs. 3 and 4 for a schematic representation) and oriented through the *coma* in the direction with a position angle of 77° on 6 October 2021 and 66° on 6 February 2022. For absolute flux calibration of the cometary spectrum, we observed the spectrophotometric standard stars BD + 25d4655 and BD + 75d325 from Oke (1990). The spectral transparency of the Earth's atmosphere for the SAO was provided by Kartasheva and Chunakova (1978). The spectrum of a He-Ne-Ar lamp was used for the wavelength calibration of the cometary spectrum. For flat-field correction of spectral images, the spectrum of the built-in LED system was used, which provides an almost uniform intensity distribution over a wide wavelength range. In addition to bias subtraction and flat-field correction, the processing of spectral images included geometrical correction along the slit, correction of the spectral line curvature, sky background subtraction, spectral sensitivity of the instrument, spectral wavelength calibration, and presentation of data with uniform scale spacing along the wavelengths. The wavelength scale calibration included automatic line identification, two-dimensional approximation of the dispersion curve by a third-order polynomial,

---

[1] https://vizier.cds.unistra.fr/viz-bin/VizieR?-source=II/336

**Table 1.**
Log of observations of comet 67P/Churyumov− Gerasimenko.

| Date[1], UT | $r^2$ [au] | $\Delta^3$ [au] | $\alpha^4$ [deg] | $\varphi^5$ [deg] | Filter/grating | $T_{exp}^6$ [s] | $N^7$ | Mode |
|---|---|---|---|---|---|---|---|---|
| The 6-m BTA | | | | | | | | |
| 021 Oct 6.954 | 1.254 | 0.475 | 47.9 | 266.7 | g-sdss | 20 | 5 | Image |
| 2021 Oct 6.956 | 1.254 | 0.475 | 47.9 | 266.7 | r-sdss | 10 | 5 | Image |
| 2021 Oct 6.974 | 1.254 | 0.475 | 47.9 | 266.7 | VPHG1200@540 | 60 | 5 | Spectra |
| 2022 Feb 6.874 | 1.675 | 0.709 | 10.5 | 133.5 | CN | 240 | 3 | Image |
| 2022 Feb 6.884 | 1.675 | 0.709 | 10.5 | 133.5 | BC | 300 | 3 | Image |
| 2022 Feb 6.895 | 1.675 | 0.709 | 10.5 | 133.5 | RC | 40 | 3 | Image |
| 2022 Feb 6.912 | 1.675 | 0.709 | 10.5 | 133.5 | VPHG1200@540 | 300 | 6 | Spectra |
| The 2-m LT | | | | | | | | |
| 2021 Oct 9.138 | 1.247 | 0.468 | 48.2 | 267.5 | r-sdss | 20 | 5 | Image |
| 2022 Feb 6.068 | 1.668 | 0.701 | 10.8 | 135.5 | r-sdss | 20 | 5 | Image |
| 2022 Feb 7.989 | 1.683 | 0.720 | 11.0 | 131.2 | r-sdss | 20 | 5 | Image |

[1] Date is the mid-cycle time.
[2] $r$ is the heliocentric.
[3] $\Delta$ geocentric distances.
[4] $\alpha$ is the phase angle of the comet.
[5] $\varphi$ is the position angle of the scattering plane.
[6] $T_{exp}$ the total exposure time.
[7] $N$ is the number of observation cycles.

quadratic smoothing of the polynomial coefficients along the slit height, and image linearization.

The application of this information to the observations of specific comets with SCORPIO-2 in spectral and photometric modes is given in the following papers: comet C/2009 P1 (Garradd) (Ivanova et al., 2017b); comet 2P/Encke (Rosenbush et al., 2020); comet 67P/Churyumov−Gerasimenko in the 2015/16 apparition (Ivanova et al., 2017a; Rosenbush et al., 2017). The summary log of observations of comet 67P/C-G, including those performed with the 6-m telescope, is given in Table 1.

*2.2. Imaging photometry at the 2-m LT*

To study the morphology of the coma and some properties of dust in comet 67P/C−G, we used photometric observations with the 2-m LT (see Table 1). The data were calibrated through a custom-built automated pipeline (Gardener et al., 2022; Giorgini et al., 1996; Lang et al., 2010; Bertin and Arnouts, 1996; Kelley and Lister, 2019). The observational circumstances for the LT observations are listed in Table 1.

**3. Spectroscopy results**

To increase the S/N ratio, we aligned and summed all the spectral images using a robust averaging algorithm. The *spectra* of comet 67P/C−G *obtained* on 6 October 2021 and 6 February 2022 consist of the strong continuum formed by the scattering of the solar light on dust grains and the emission spectrum. Most of the emission lines (sky lines) observed in the red range of the spectrum come from the Earth's atmosphere. The strongest cometary emissions are from CN ($B^2\Sigma^+ - X^2\Sigma^+$, $\Delta \nu = 0$) at about $\lambda 3883$ Å and $C_2$ ($d^3\Pi_g - a^3\Pi_u$, $\Delta \nu = 0$) Swan band at $\lambda 5165$ Å. These emission bands cover *the entire* height of the slit. Because of this, the night sky spectrum was taken from the spectrum of the standard star and then carefully subtracted from the cometary spectrum.

The measured fluxes from comet 67P/C−G at a given wavelength were obtained using a rectangular diaphragm with a half-width of about 10,000 km and a length equal to the width of the slit centered at maximum brightness. The flux error depends on the signal accumulation time, observation conditions, the wavelength range, and the distance from the optocenter, that is, the projected cometocentric distance $\rho$. The relative error $\varepsilon = F_\lambda^{err}/F_\lambda$, where $F_\lambda$ and $F_\lambda^{err}$ are, respectively, the total flux density and its error. They are measured within the aperture of 10,000 km at different wavelengths and are displayed in Fig. 1. Since the sensitivity of the CCD matrix used is low in the blue range of the spectrum, the errors in the measured radiation fluxes are quite large, as shown in this figure.

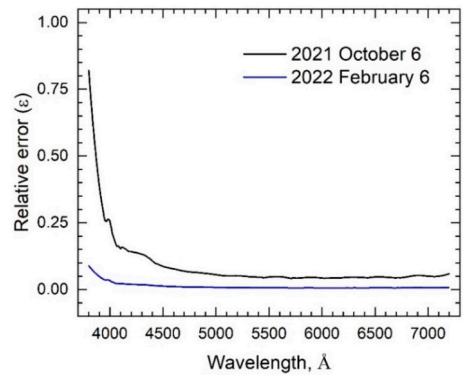

**Fig. 1.** Relative errors in the measured fluxes from comet 67P/Churyumov−Gerasimenko in the aperture of 10,000 km on 6 October 2021 (black line) and 6 February 2022 (blue line). (For interpretation of the references to color in this figure legend, the reader is referred to the web version of this article.)

Fig. 2 shows the stages of the reduction of the cometary spectrum. To separate the continuum signal from the gaseous emissions, we applied the method described by Ivanova et al. (2018). For this, we used the high-resolution solar spectrum by Neckel and Labs (1984), which was transformed to the resolution of our cometary spectra using convolution with the instrumental profile and normalization to the flux of the comet. We compared the convolved solar spectrum and the spectrum of comet 67P/C−G in Fig. 2a. Then we calculated a polynomial from the fluxes measured in both solar and cometary spectra in the continuum windows ($\lambda$3800–3830, 4135–4165, 4230–4260, 4417–4483, 5232–5288, 5760–5820, 6150–6200, 6420–6460, and 7099–7157 Å), using the IDL software *robust_poly_fit.pro* which provided an outlier-resistant polynomial fit. The degree of the fitting polynomial was chosen according to the minimum absolute deviations and is shown for each observation date in Fig. 2b. Multiplying the solar spectrum by these curves, we obtained the continuum spectra, which were then subtracted from the original spectrum of the comet for each date. Thus, we obtained the pure emission spectra of comet 67P/C−G for both observation dates which are shown together with individual emission bands in Fig. 2c. The CN ($B^2\Sigma^+ - X^2\Sigma^+$), $C_3$ ($\tilde{A}^1\Pi_u - \tilde{X}^1\Sigma_g^+$), $C_2$ ($d^3\Pi_g - a^3\Pi_u$), and $NH_2(\tilde{A}^2A_1 -$

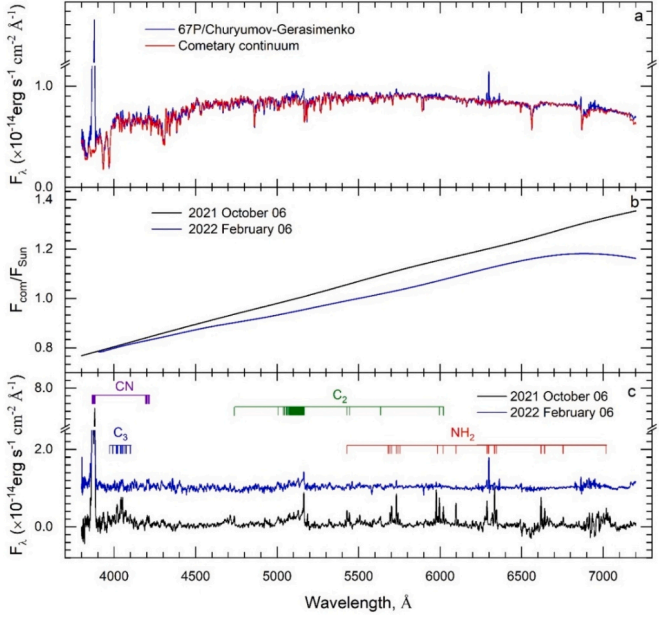

**Fig. 2.** Reduction of the spectra of comet 67P/Churyumov–Gerasimenko obtained on 6 October 2021 and 6 February 2022. Panel (a) shows the spectrum of the comet (blue line) for 6 October 2021 with the scaled solar spectrum (red line); (b) is the normalized spectral dependence of the dust reflectivity; (c) shows two emission spectra of the comet, shifted along the Y axis for different observation dates. (For interpretation of the references to color in this figure legend, the reader is referred to the web version of this article.)

$\widetilde{X}^2B_1$) bands are identified in the emission spectrum.

For the calculation of the production rate of CN, $C_3$, $C_2$, and $NH_2$ molecules, we used the model by Haser (1957). Input parameters for this model, namely, the fluorescence efficiency (the g-factor), characteristic scale lengths for the parent $l_p$ and daughter $l_d$ molecules, and power-law index $n$, dependent on the heliocentric distance according to $r^{-n}$, were taken from (Langland-Shula and Smith, 2011). The g-factor for the CN ($\Delta v = 0$) band and heliocentric velocity of the comet depend on the heliocentric distance. Schleicher (2010) published the tables of the g-factor values for different heliocentric distances and velocities, which we used in our calculations. To calculate the gas production rate of $NH_2(0,7,0)$, the fluorescence efficiency factor was taken from Kawakita et al. (2001). In Table 2, we present the parameters used for the Haser model. The Haser formula is defined for a circular aperture and, hence, should be modified for the case of a rectangular aperture of the spectrograph slit. For this, we extracted a useful signal of the comet from the image of a rectangular slit which corresponds to the flux through a circular diaphragm, the center of which is located on the comet optocenter. Using the parameters given in Table 2, we computed the production rates of CN, $C_3$, $C_2$, and $NH_2$ molecules in comet 67P/C–G which are shown in Table 3. Since the $C_3$ ($\tilde{A}^1\Pi_u - \widetilde{X}^1\Sigma_g^+$) band is practically no different from the noise background, we assume that this is the upper limit of the production rate value for observations in 2022.

Based on the composition determined by the ratio of production rates of different species, A'Hearn et al. (1995) divided comets into two classes according to their gas production rates: typical comets and carbon-chains depleted comets. The depleted comets are characterized by the $\log[Q(C_2/CN)] < -0.18$ ratio, while the mean value of this ratio for typical comets is 0.06. We do not analyze the $\log[Q(C_3/CN)]$ ratio because no significant $C_3$ was detected. According to the Schleicher (2008) comet database, the average relative molecular abundance for depleted comets is $\log[Q(C_2/CN)] = -0.61 \pm 0.35$ which coincides with our value in Table 3. In addition, Langland-Shula and Smith (2011) found that the mean $\log[Q(C_2/CN)]$ ratio for a subset of the Jupiter Family comets is $-0.7 \pm 0.2$, which is close to our calculated value within the uncertainty range. Thus, comet 67P/C–G can be classified as a $C_2$-depleted comet. It should be noted that large differences in relative abundances between the summer and winter hemispheres were found by the ROSINA instruments (Le Roy et al., 2015). So, the comet 67P/C–G can be carbon-chain depleted for the summer hemisphere, whereas, for the winter hemisphere, the comet would be carbon-chain normal. However, the summer hemisphere contributes most significantly to ground-based observations of the comet to compare to space mission observations. In our case, the subsolar latitudes on October 6, 2021, and February 6, 2022, were $-32.6°$ and $-29.5°$, respectively, indicating illumination of the summer hemisphere. On the other hand, the heliocentric dependence of the $C_2/CN$ production rate ratio might also be responsible for those differences.

We also explored the behavior of reflectivity as a function of wavelength in the visible region. The slope of the spectrum $S(\lambda)$ along the dispersion, which is reddening, is defined as the ratio of the cometary spectrum $F_{com}(\lambda)$ to the scaled solar spectrum $F_{sun}(\lambda)$: $S(\lambda) = F_{com}(\lambda)/F_{sun}(\lambda)$. The behavior of $S(\lambda)$ is shown in Fig. 2b. For comparison of the normalized spectral gradient of the reflectivity $S'(\lambda_1, \lambda_2)$ (in % per 1000 Å) found for comet 67P/C–G with that for other comets, we used the same expression (Jewitt and Meech, 1986) as in Kwon et al., 2022:

$$S'(\lambda_1, \lambda_2) = \frac{1}{S_{mean}} \frac{S(\lambda_2) - S(\lambda_1)}{\lambda_2 - \lambda_1}, \quad (1)$$

where $S(\lambda_1)$ and $S(\lambda_2)$ are the continuum flux measurements performed at wavelengths $\lambda_1$ and $\lambda_2$ with the condition $\lambda_2 > \lambda_1$. The result obtained shows that the dust reflectivity increases with increasing wavelength. Since two sets of filters were used for observations, broadband g-sdss and r-sdss ($\lambda$4770–6231 Å) and cometary narrowband BC and RC ($\lambda$4450–6839 Å), we determined $S'$ in these two domains. As a result of observations on 6 October 2021, we obtained the value of the normalized reflectivity gradient $(16.1 \pm 4.4)$ % 1000 Å$^{-1}$ for the broadband range and $(15.7 \pm 3.8)$ % 1000 Å$^{-1}$ for the narrowband range. For observations on 6 February 2022, $(15.2 \pm 0.5)$ % 1000 Å$^{-1}$ for the first domain and $(13.3 \pm 0.4)$ % 1000 Å$^{-1}$ for the second domain.

**Table 3.**
Gas production rates in comet 67P/Churyumov–Gerasimenko obtained from spectral observations during the 2021/22 apparition.

| Date, UT | | r [au] | Q [mol s$^{-1}$] | | | | Production rate ratio logQ | | |
|---|---|---|---|---|---|---|---|---|---|
| | | | CN | $C_3$ | $C_2$ | $NH_2$ | ($C_3$/CN) | ($C_2$/CN) | ($NH_2$/CN) |
| 2021 Oct 6 | | 1.25 | (8.8 ± 6.6) × 10$^{24}$ | (3.8 ± 1.1) × 10$^{23}$ | (6.1 ± 0.3) × 10$^{24}$ | (1.2 ± 0.03) × 10$^{24}$ | −1.3 ± 0.5 | −0.2 ± 0.4 | −0.8 ± 0.3 |
| 2022 Feb 6 | | 1.68 | (1.7 ± 0.2) × 10$^{24}$ | (0.6 ± 3) × 10$^{22}$ | (4.3 ± 0.8) × 10$^{23}$ | (3 ± 0.6) × 10$^{23}$ | −2.4 ± 2.1 | −0.6 ± 0.1 | −0.7 ± 0.1 |

**Table 2.**
Model parameters reduced to the heliocentric distance 1 au.

| Molecule | g-factor [W mol$^{-1}$] | $L_p \times 10^4$ [km] | $l_d \times 10^4$ [km] | n |
|---|---|---|---|---|
| CN ($\Delta v = 0$), $\lambda$3883 Å | $4.4 \times 10^{-20}$ | 0.5 | 5.0 | 2 |
| $C_3$, $\lambda$4050 Å | $1.0 \times 10^{-19}$ | 0.2 | 3.4 | 2 |
| $C_2$ ($\Delta v = 0$), $\lambda$5165 Å | $4.5 \times 10^{-20}$ | 0.7 | 1.1 | 2 |
| $NH_2$ (0,7,0), $\lambda$6621 Å | $2.29 \times 10^{-21}$ | 0.3 | 2.1 | 2 |

## 4. Morphology of dust and gas coma

To improve our understanding of the spatial structures in the coma of comet 67P/C−G, which arise from the anisotropic ejection of gas and dust from the nucleus, we used the images taken with the *g*-sdss and *r*-sdss filters on 6 October 2021, −31 days before passing perihelion, and with narrowband comet HB filters (Farnham et al., 2000), continuum *BC* (λ4450/62 Å) and *RC* (λ6839/96 Å) filters and *CN* (λ3870/58 Å) emission filter, on 6 February 2022, when the comet was +96 days after perihelion. We used both the wideband *r*-sdss and the narrowband *BC* and *RC* as a proxy for dust. To enhance low-*contrast* structures in the cometary images, we applied an *enhancement* technique to all images, namely the Larson and Sekanina (1984) rotational gradient technique and median normalization. These digital filters allow us to reveal subtle morphological details in the comet that cannot be detected in the raw images.

### 4.1. Large-scale morphology of dust coma

Fig. 3 (a, d) illustrates the stacked images of comet 67P/C−G taken with the *g*-sdss and *r*-sdss filters on 6 October 2021. The projected direction of the Sun is indicated at the top of the images. The comet shows a dust tail in the antisolar direction at the position angle $PA_{\text{tail}} \approx 266°$, measured counterclockwise from the north through the east, and, as isophotes demonstrate, an asymmetric coma in the solar direction. The *r*-sdss image exhibits a more extended coma and dust tail than those obtained in the *g*-sdss filter. To highlight low-contrast structures in the images, we applied the rotational gradient method developed by Larson and Sekanina (1984) to all individual images of the comet and the composite image to avoid introducing false structures that may occur after applying the enhancement technique. Enhanced images obtained in both the *g*-sdss and *r*-sdss filters are displayed in the right panels (c) and (f). These images show a jet-like structure directed toward the Sun and oriented near position angle $PA_{\text{jet}} \approx 127°$. A similar structure of the coma, namely the presence of the sunward jet and dust tail, was also detected in the image of comet 67P/C−G obtained on 11 November 2021 by Biver et al. (2023). The jet and tail, projected onto the sky, appear as a radial dust outflow from the nucleus. However, the jet exhibits a positive curvature (counter-clockwise) with increasing distance from the nucleus, while the tail has a slight negative curvature (clockwise). We found no traces of gas or plasma tail in our images.

As shown in Fig. 4, the structure of the dust coma of comet 67P/C−G changed significantly during post-perihelion (+96 days) observations on 6 February 2022, in comparison with pre-perihelion images of the comet. Panels (a) and (d) are the direct images of the comet in the narrowband continuum *BC* and *RC* filters, whereas (b) and (e) are images with superimposed relative isophotes. These images show an asymmetric dust coma, elongated approximately perpendicular to the solar-antisolar direction. However, the most notable feature of these images is the bright linear structure observed in the sunward direction at a position angle of ∼283°, which is projected against the background of the extended coma. It is especially clearly visible in the RC filter, indicating a dust structure because a sunward gas tail cannot be observed. As Fig. 4 shows, this linear strip extends in the direction of the Sun, but does not emerge exactly from the optical center of the coma: it is displaced in the north direction. It can be assumed that this strip-like structure corresponds to one edge of a side-on cone swept out by the rotating jet and therefore is very similar to the neckline structure (see details in Section 4.3). Taking into account that the position angle of the extended radius vector of the comet is $PA_{RV} = 134°$, this structure cannot be a gas or plasma tail.

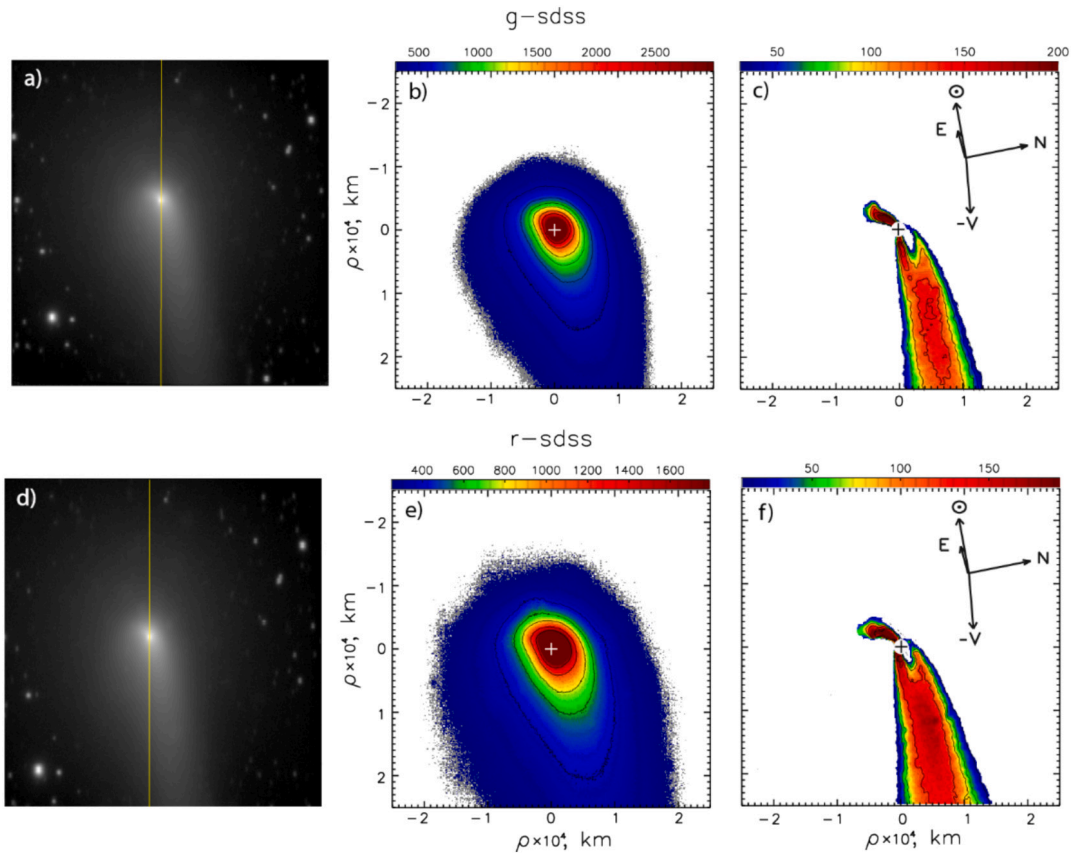

**Fig. 3.** Co-added images of comet 67P/Churyumov−Gerasimenko obtained at the 6-m telescope through the *g*-sdss and *r*-sdss filters on 6 October 2021. Panels (a) and (d) show the raw CCD images of the comet and the position of the slit (yellow line) during the spectral observations; (b) and (e) are the images with relative isophotes differing by a factor of $\sqrt{2}$ in intensity; (c) and (f) are the images to which a rotational gradient technique of Larson and Sekanina (1984) is applied. The images are color-enhanced to show contrast. The location of the optical center of the comet is marked with a cross. The arrows indicate the projected direction to the Sun (⊙), the north (N), the east (E), and the negative velocity vector of the comet in projection onto the sky plane (−V). Negative and positive distances indicate sunward and anti-sunward directions, respectively. (For interpretation of the references to color in this figure legend, the reader is referred to the web version of this article.)

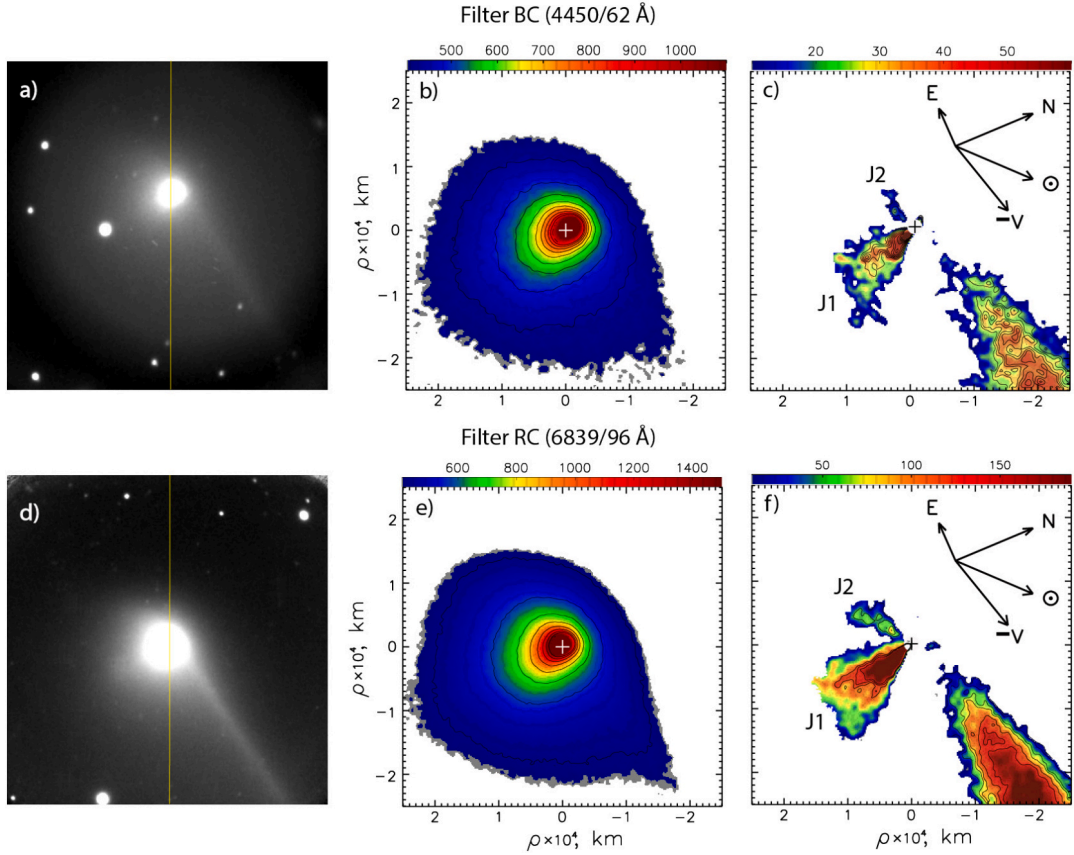

**Fig. 4.** Same as in Fig. 3, but images of comet 67P/Churyumov–Gerasimenko were taken on 6 February 2022 at the 6-m telescope with the narrowband cometary filters BC ($\lambda$4450/62 Å) and RC ($\lambda$6839/96 Å). In panels (a) and (d), the bright linear structure toward the Sun is presumed to be a neckline structure. J1 and J2 are the jet-like structures. The rest of the designations are the same as in Fig. 3.

To enhance the other possible structures in the coma, we have applied a rotational gradient technique of Larson and Sekanina (1984) to the images taken on 6 February 2022. The processed images (c) and (f) in Fig. 4, beside the neckline structure, clearly show two jet-like structures designated J1 and J2. One of these structures, J1, is approximately perpendicular to the solar-antisolar direction at $PA_{J1} \approx 193°$, and the other, J2, is directed approximately in the antisolar direction at $PA_{J2} \approx 133°$. On February 6.89, 2022 UT, the position angle of the neckline structure was approximately 283°, while the jet-like structure J2 was observed at around 133°. The difference between these angles is about 150°, which deviates significantly from 180°. This indicates that the jet-like structure J2 cannot be a direct extension of the neckline structure. Therefore, we can reasonably conclude that there is no evident connection between the two phenomena. However, it is not possible to separate them, since the neckline structure cannot be traced further beyond the nucleus. On the other hand, it is possible that the jet-like structure in the antisolar direction is caused by both the overlap of the 180° continuation of the neckline in the outer coma and contamination from large particles released during several previous orbits (see, for example, Fig. 3 in Moreno et al., 2017).

### 4.2. Numerical simulation of jets formed by the active areas on the nucleus

The geometric model of jet formation developed by Kleshchonok and Sierks (2025) allows us to trace how the jet structure observed in comet 67P/C–G arises. In this model, the configuration of jets near the nucleus is determined by the location of the observer, the Sun, and the comet, as well as the axis of rotation of the nucleus. The model does not consider a dispersion of the sizes and velocities of the jet particles and their acceleration by solar radiation, however, it allows for quickly getting results for many combinations of parameters to obtain the best fit between observed and model jets, and also to trace the evolution of the jet structure over a long period. Using estimates of the rotation period of the comet determined earlier by Mottola et al. (2014) and Sierks et al. (2015) for the two previous cometary apparitions in 2009 and 2015 and implying a decrease of the rotation period by 0.36 h during the perihelion passage, we adopted the comet's period 12.05 h for the model calculations (Fig. 5).

Calculations show that for the 6 October 2021 observation, it is impossible to select acceptable parameters for the jet formed from the optocenter. However, a good agreement between the model jet and the observed image is obtained if the optocenter does not coincide with the nucleus. In Fig. 5, each blue line shows the position of particles ejected simultaneously at a constant velocity. The time interval between adjacent lines is 0.01 days. As a result of the simulation, we have obtained the following parameters for the model jet: the active region is located on the nucleus at latitude $\varphi = -70° \pm 4°$, an opening angle of the jet is $20° \pm 6°$, and the particle emission velocity in the direction of the jet is $0.31 \pm 0.04$ km s$^{-1}$. Due to the conditions of the projection of the jet onto the sky plane, there is a visible overlap (a loop, see the inset in

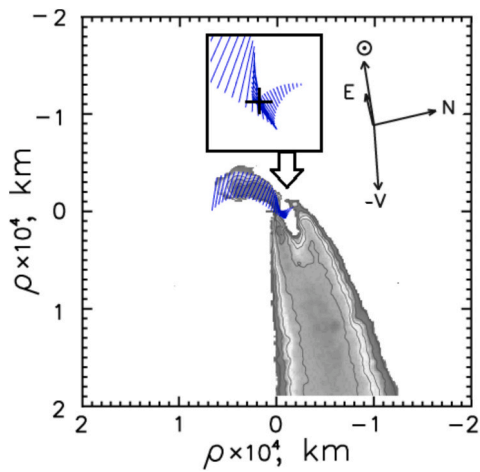

**Fig. 5.** Simulation of the jet observed on 6 October 2021 using a geometric model by Kleshchonok and Sierks (2025). Blue lines show the position of particles ejected simultaneously from the nucleus at a constant velocity, and the time interval between neighboring lines is 0.01 days. The inset shows in an enlarged scale a visible overlap (a loop) of jet particles ejected at different times. The cross and arrows are the same as shown and described in Fig. 3. (For interpretation of the references to color in this figure legend, the reader is referred to the web version of this article.)

Fig. 5) of jet particles ejected at different times. As a result of such superposition, a local increase in brightness is observed, which leads to a visible displacement of the optocenter from the actual position of the nucleus.

The enhanced images taken on 6 February 2022, show two jets, J1 and J2, located symmetrically relative to the axis of rotation of the nucleus (Fig. 6). The separation of jets in the image was achieved using a digital rotational gradient filter, which highlights the radial components of jets but cannot highlight broader jets azimuthally oriented within the coma. Therefore, only the radial parts of the jets are visible in the filtered image of the coma. The symmetry suggests that both jets originate from the same active source on the rotating nucleus. In this case, part of the jet should also be located between the radial jets visible in the image. This is confirmed by simulation using a geometric model: the segments between the two selected jets are predominantly aligned in the azimuthal direction. With this model, we determined that this active area is located at a latitude of $\varphi = -58° \pm 5°$ with an opening angle of the jet of $26° \pm 8°$. The velocity of particles in the jet ejected from the active area is estimated at $0.30 \pm 0.05$ km s$^{-1}$. It should be noted that a similar formation of two observed jets from the same area was also revealed in comet 2P/Encke (Rosenbush et al., 2020).

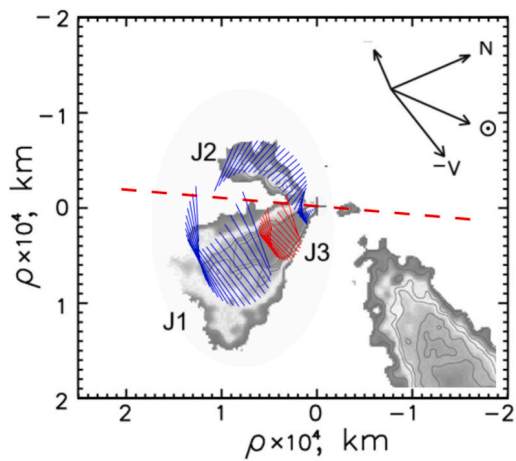

**Fig. 6.** The geometric model illustrates the formation of jets from the nucleus of comet 67P/Churyumov−Gerasimenko on 6 February 2022 in the RC filter. The model jets J1 and J2 emanating from the first active area are depicted by a set of blue lines, and the J3 jet, formed by the second active area, is shown by red lines. The time interval between the blue and red lines is 0.01 days. The red dash-dotted line shows the projection of the nucleus rotation axis onto the celestial plane. The cross and arrows are the same as shown and described in Fig. 3. (For interpretation of the references to color in this figure legend, the reader is referred to the web version of this article.)

The velocity value obtained in comet 67P/C-G is close to those of dust particles obtained for other comets. For example, Jorda et al. (1994) determined an expansion velocity of the dust of $v = 0.3$ km s$^{-1}$ in the jet of comet P/Swift-Tuttle 1992 at a heliocentric distance $r \approx 1$ au. According to Warell et al. (1997), the projection onto the plane of the sky of the velocity of particles forming shells in comet C/1995 O1 (*Hale-Bopp*) was 0.41 km s$^{-1}$. Stansberry et al. (2004) determined the particle velocity of 0.5 km s$^{-1}$ at a distance of $r \approx 6$ au in comet 29P/Schwassmann-Wachmann 1. Contrary, Garcia et al. (2024) obtained $v = 0.18$ km s$^{-1}$ at $r = 2.3$–2.6 au for comet C/2017 K2 (PanStarrs), explaining such a low dust velocity by the presence of predominantly large particles in the coma.

As Fig. 4 shows, the jets differ very much in intensity: jet J1 appears much brighter than the J2 jet. This difference occurs because the J1 jet forms at the moment of maximum illumination of the active region by the Sun, which leads to a more intense ejection of gas and dust from this region. And vice versa, the J2 jet forms when the angle of incidence of sunlight on the active region reaches its maximum, which leads to a decrease in its activity and, accordingly, to a weakening of the J2 jet. If our assumption is correct and the same active region generates both jets, we should observe an interruption of the J1 jet due to the nucleus rotation that occurs during the formation of the J2 jet. Since such a gap is not observed in the J1 jet, we may assume that there is another active source on the cometary nucleus, from which escaping particles fill the gap in the J1 jet, while the first active area forms the J2 jet. According to our calculations within the framework of the geometric model, the second active area is located at latitude $\varphi = -53° \pm 10°$ and separated from the first one by a difference in longitude of $150° \pm 20°$. This area forms jet J3 (red lines in Fig. 6) which overlaps jet J1 and fills the gap that is created by the activity of the first active area.

A distinctive feature of the second active region is that it is active only during part of the rotation period, namely when it is illuminated by the Sun. This can be explained by the peculiarities of the comet's nucleus relief. For example, if an active region is located in a depression, it will be illuminated by the Sun for only part of the rotation period, for a shorter duration than if the nucleus were spherical. According to Fornasier et al. (2019), there are two regions on the surface of the nucleus of comet 67P/C-G which satisfy all the conditions specified by the geometric model. The first area is located near the boundary of the Khonsu and Bes formations on a convex part of the nucleus, on the side of the larger lobe, thus receiving sufficient sunlight. The second area is located near the border of Anhur and Bes and lies in a depression, in the "neck" between the two nucleus lobes, receiving less sunlight than the first area. According to Fornasier et al. (2019), these regions exhibit a very high activity after perihelion passage. Although other combinations of active areas are possible, the most plausible locations appear to be near the boundaries of Khonsu-Bes and Ankhur-Bes.

The location of the active areas calculated using the geometric model is shown in Fig. 7. For calculations, we used the shape model of comet 67P/C−G "RO-C-MULTI-5-67P-SHAPE-V2.0" available at the site. In Fig. 7, we show the 3D model of the nucleus as observed on 6 October 2021 (panel (a)) and on 6 February 2022 (panel (b)). The simulated locations of active jet-forming areas are highlighted in color. The black arrow indicates the orientation of the nucleus rotation axis. The intensity of the white shading on the nucleus corresponds to the degree of local illumination of the nucleus by the Sun.

As seen in Fig. 7a, on 6 October 2021, the region around latitude $-70°$ is well illuminated by the Sun, supporting high activity in this part of the nucleus and forming a powerful jet. On 6 February 2022 (panel b), the active area at around latitude −58° (in blue) is also well illuminated by the Sun supporting its high activity. In contrast, the region at latitude −53° (marked in red) receives less sunlight over a full rotation period due to its location in a depression. This can be seen in panel (b), where the possible active area (in red) is poorly illuminated and is in shadow, while neighboring higher elevation regions are still well-lit by the Sun. As a result, the geometric model predicts that the active region producing the J3 jet will remain active for a short fraction of the nucleus's rotation period due to its location.

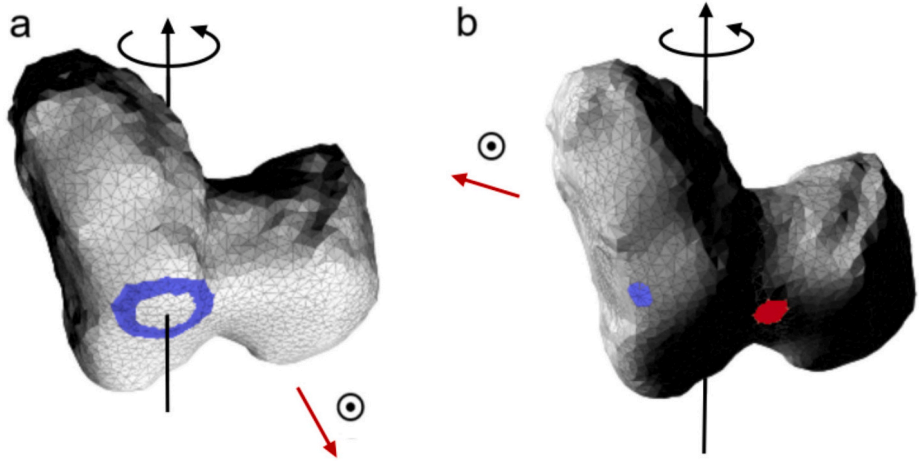

**Fig. 7.** The 3D model of the nucleus of comet 67P/Churyumov-Gerasimenko. Panel (a) shows the nucleus as observed on 6 October 2021, and the assumed area of the formation of the jet (see Fig. 3) is outlined by a blue ring. Panel (b) displays the nucleus on 6 February 2022 with simulated locations of the active areas generating the J1 (blue spot) and J3 (red spot) jets (see Fig. 4). The intensity of the white color corresponds to the degree of local illumination of the nucleus by the Sun. The black arrow indicates the orientation of the nucleus rotation axis, and the red arrow points toward the Sun. (For interpretation of the references to color in this figure legend, the reader is referred to the web version of this article.)

We compared the obtained coordinates of the active areas with the corresponding data determined by Boehnhardt et al. (2024) for dates close to our observation dates. On 6 October 2021, we identified an active region at latitude − 70° ± 4°. On nearby dates, on 2 and 11 October 2021, Boehnhardt et al. established the presence of jets from three active areas: A: +40° ± 5°; C: −50° ± 5°; D: −70° ± 5°. Our results are consistent with the D area. According to modeling jets observed on 6 February 2022, we identified two active regions at latitudes −58° ± 5° and − 53° ± 10° separated by a longitude of 150° ± 20°, while Boehnhardt et al. on 18 February 2022 found the jets from areas A: +40° ± 5° and C: −50° ± 5°. The positions of the −58° ± 5° and − 53° ± 10° are close in latitude to the area C −50° ± 5° within the error range. The authors note that region C is observed as the boundaries of a cone with an axis that coincides with the position of the nucleus rotation axis. Therefore, we can state that our determination of the jet parameters derived from the 6 February 2022 images is consistent with the results presented by Boehnhardt et al. (2024).

### 4.3. The neckline structure

Visible in the image taken on 6 February 2022 the bright linear-like structure elongated in the sunward direction (Fig. 4) strongly resembles a neckline structure (Kimura and Liu, 1975). The neckline was earlier observed in comet 67P/C−G in optical and near-infrared domains and analyzed by many authors, among whom Fulle (2004), Agarwal et al. (2006, 2007a, 2007b, 2010), Ishiguro (2008), Kelley et al. (2008, 2009), Fulle et al. (2010), Lara et al. (2011a, 2011b), Moreno et al. (2017). The neckline structures of about the same nature as those of comet 67P/C−G have been observed in many comets, e.g., comets C/1956 R1 (Arend-Roland) (Kimura and Liu, 1975), C/1969 Y1 (Bennett) (Pansecchi et al., 1987; Fulle and Sedmak, 1988), 1P/1982 U1 (Halley) (Sekanina et al., 1987; Cremonese and Fulle, 1989), C/1995 O1 (Hale-Bopp) (Boehnhardt, 2003), and others. For completeness, it is worth mentioning that a well-defined neckline was also observed in the short-period comet 22P/Kopff (Moreno et al., 2012), further confirming that such structures are not limited to long-period comets.

According to Fulle (2004), it is likely that earlier observed anti-tails projected onto the sky plane were, in fact, neckline structures. However, it should be noted that other authors, like Agarwal et al. (2007b), believe that the neckline structure is not a part of the comet, but is an effect of perspective caused by the Earth moving through the comet's orbital plane.

The parameters of several neckline structures in comet 67P/G-C are presented in Table 4, showing they occur post-perihelion and within a specific orbital window. The lifetime of necklines is shorter than the comet's orbital period, as planetary tides and solar radiation gradually disperse the dust. The neckline's position angle varies only slightly

**Table 4.**
The observed characteristics of neckline structures in comet 67P/Churyumov−Gerasimenko.

| Date, UT[1] | $N_d$[2] | $r$[3] [au] | $\alpha$[4] [deg] | $PA_l$[5] [deg] | $PA_s$[6] [deg] | $PA_{NL}$[7] [deg] | Reference |
|---|---|---|---|---|---|---|---|
| 2003 Mar 27 | 221 | 2.61 | 9.7 | −3.0 | 313.2 | 293.5 | Lara et al., 2005 |
| 2003 Mar 5 | 258 | 2.87 | 18.2 | −0.5 | 296.3 | 294.5 | Agarwal et al., 2006 |
| 2015 Dec 19 | 128 | 1.92 | 30.8 | −2.4 | 113.9 | 302.1 | Moreno et al., 2017 |
| 2016 Mar 8 | 208 | 2.53 | 4.4 | −4.0 | 310.8 | – | Moreno et al., 2017 |
| 2016 Mar 29 | 229 | 2.68 | 8.4 | −2.7 | 315.3 | 296.0 | Moreno et al., 2017 |
| 2022 Feb 6 | 96 | 1.67 | 10.1 | −5.4 | 315.5 | 284.1 | This work |

[1] Date is the mid-cycle time.
[2] $N_d$ is the number of days after perihelion.
[3] $r$ is the heliocentric.
[4] $\alpha$ is the phase angle of the comet.
[5] $P_l$ is the angle between the observer and target orbital plane.
[6] $PA_s$ is the position angle of the extended Sun-to-target radius vector.
[7] $PA_{NL}$ is the position angle of the neckline structure.

within this window and is sensitive to the comet–Sun–Earth geometry.

Images from 19 December 2015 and 29 March 2016 show similar coma structures despite differing Sun projection angles. At opposition, the Earth crosses the Sun–comet line, causing the tail's projection to appear reversed, creating a so-called "anti-tail" pointed sunward. This is a projection effect; the tail's actual direction changes little.

On 6 February 2022, the comet was imaged shortly after passing the opposition point. Although the true direction toward the Sun in space remains virtually unchanged, its projection onto the sky reverses (Fig. 9b), creating the appearance that the entire dust tail is directed toward the Sun. A similar situation occurred during the observation of comet 67P/G-C on 29 March 2016, as seen in Fig. 9a.

The conditions for observing the neckline must be such that we can see the dust being ejected when the comet's true anomaly differs by 180° from its value at the time of observation (see Table 4). For such a neckline to be observed, a significant dust release would have to occur around 15 April 2021. But at that time, the comet was far from the Sun, at a heliocentric distance of 2.49 au, and therefore a significant dust emission is unlikely. The maximum dust emission occurred most likely near the perihelion date of 2 November 2021, more than 3 months before our observations on February 6, 2022. Thus, it can be assumed that the observed fine structure is primarily formed due to the ejection of large dust particles near the comet's perihelion. This is further confirmed by modeling the motion of dust particles in the coma using the Finston and Probstein (1968) model [Comet Toolbox]. In Fig. 10 (on the right), we demonstrated the existence of a long narrow dust tail at the same position angle as the observed structure in the comet image. The figure displays synchrones and syndynes, and the numbers indicate the lifetime of the synchrones in days. Synchrones describe the positions of particles with different β emitted at a given time for the following values of $\beta$: 0.05, 0.1, 0.15, 0.2. In Fig. 10, the syndynes are not labeled due to their spatial proximity.

We interpret the visible sunward tail 6 February 2022 (Fig. 10, left image), as a projection of small-particle dust, with the neckline structure formed by larger particles released near perihelion superimposed upon it.

### 4.4. The CN coma

A'Hearn et al. (1986) were first to detect cyanogen (CN) jets in comet 1P/Halley. At that time, it was believed that various structures could only be observed in the dust coma, but they could not be seen in the emission bands of the typical neutral species. The reason for this could be the high velocity of expansion of both neutral parent molecules in expanding halos and daughter species produced by dissociations occurring far from the place of origin of the gas. According to A'Hearn et al., the CN images of comet Halley showed prominent jets extending to >60,000 km from the nucleus in different directions and lasting several weeks. Due to the rotation of the nucleus, the jets were curved or had a spiral shape. At the same time, there were no dust jets in the images of the coma taken in the continuum. Subsequently, CN jets were observed by different authors in many comets, e.g. comets 1P/Halley (Cosmovici et al., 1988), 109P/Swift−Tuttle (Boehnhardt and Birkle, 1994), C/2004 Q2 (Machholz) (Farnham et al., 2007), 9P/Tempel 1 (Boehnhardt et al., 2007), 8P/Tuttle (Waniak et al., 2009), 2P/Encke (Ihalawela et al., 2011), 103P/Hartley 2 (Samarasinha et al., 2011; Meech et al., 2011; Lara et al., 2011a, 2011b; Waniak et al., 2012), 45P/Honda−Mrkos−Pajdušáková (Springmann et al., 2022), and others.

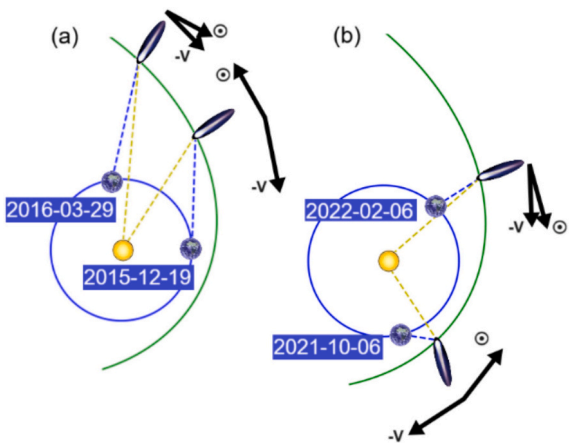

**Fig. 9.** Sketch of the observation geometry for the 2015/16 and 2021/2022 apparitions. Panel (a) shows the configuration of the Sun, Earth, and comet 67P/Churyumov-Gerasimenko during observations made by Moreno et al. (2017) in the 2015/2016 apparition before the opposition. For comparison, panel (b) shows the same configuration of the Sun, Earth, and the comet during our observations in the 2021/2022 apparition after the opposition.

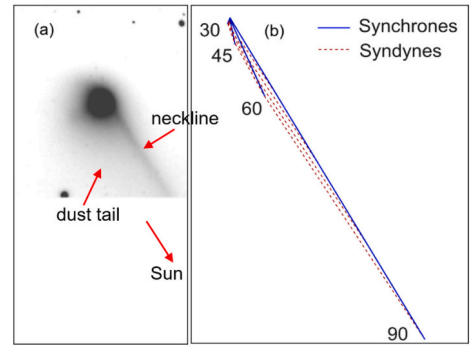

**Fig. 10.** Comparison of the observed neckline in comet 67P/Churyumov-Gerasimenko with synchrones calculated using the method by Finston and Probstein (1968).

We found a completely different structure of the CN coma in comet 67P/C−G observed through the CN filter ($\lambda$3870/58 Å). This filter minimizes the contamination from the continuum and therefore a dust image was not subtracted. Our spectral observations on February 6, 2022, revealed high line intensities of CN, suggesting that the coma image obtained through the CN filter primarily reflects the distribution of CN molecules, with only a minor contribution from the dust continuum. Panel (a) in Fig. 11 displays the direct CN image of the comet, whereas panel (b) shows the same image with superimposed isophotes. Preliminary inspection of these images revealed an asymmetric coma with a bright compact spot around the nucleus, up to ~5000 km, that may indicate a rapid release of a CN parent material. Panel (c) shows the CN image that was enhanced by dividing by the median. As can be seen, neither the tail nor the jets, which are visible in the continuum BC and RC images of comet 67P/C−G (see Fig. 4), are observed in the CN image. It should be noted that some comets exhibit gas features consistent with dust structures, while others show no correlation. On the contrary, there are comets with distinct gas features, although their dust structures are very weak or completely absent. In the case of comet 67P/C−G, the CN coma seems to be approximately elongated in the direction between the two dust jets. It makes sense if this is associated with strong activity of one of the hemispheres of the nucleus. This is in agreement with observations taken on the same night using the MUSE instrument at the ESO VLT, which also indicate dust and CN features aligned in broadly the same direction, to the south of the nucleus (Murphy et al., 2025).

Since the phase angle of the comet at the time of observation was quite small, $\alpha = 10.5°$, nearly circular isophotes should have been observed. However, the elongated shape of the observed CN coma and, therefore, the isophotes significantly deviate from circular symmetry, and their conventional axis of symmetry, located at a position angle of ~162°, shows a noticeable shift from the direction opposite to the Sun. As shown in Fig. 11, the position angle of the prolonged radius vector

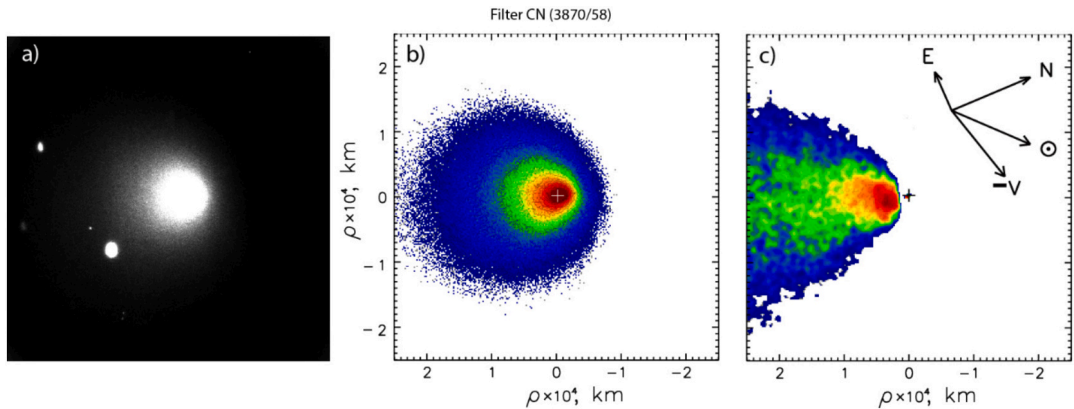

**Fig. 11.** Image of the coma of comet 67P/Churyumov–Gerasimenko taken with the narrowband cometary *CN* (*λ*3870/58 Å) filter at the 6-m telescope on 6 February 2022. Panel (a) is a direct CN image of the comet derived after bias and flatfield corrections and sky background subtraction; (b) is the CN image with relative isophotes differing by a factor of $\sqrt{2}$ in intensity; (c) is a digitally processed image that has been normalized by dividing by the median. The images are color-enhanced to show contrast. The negative distance is in the solar direction, and the positive distance is in the anti-solar direction. Cross and arrows are the same as shown and described in Fig. 3.

(*direction* away from the *Sun*) is 133.5°. This deviation can be explained by suggesting that the dust jets visible in the images obtained with continuum filters are an additional source of CN radicals in the coma. This was reported by Hänni et al., 2020 during a study of the CN coma near the nucleus of comet 67P/C–G.

Fig. 12 illustrates the model jets (blue lines), and the rotation axis of the nucleus indicated by a red dotted line. As shown in the figure, the Sun heats the western part of the nucleus more intensely. Therefore, the more active jet is also located on the west side of the nucleus. If the jets are indeed an additional source of CN, then the western part of the CN coma should be more developed. In this case, the conditional axis of the CN coma should also be shifted to the west relative to the axis of rotation and a correlation should be observed between the CN coma and the structure of the jets. All these features are observed in the image of the CN coma.

Under the influence of sunlight, CN radicals acquire additional acceleration in the direction opposite to the Sun. This occurs because when a photon is absorbed, it transfers momentum to the CN radical in the direction away from the Sun, causing the electrons in the radical to be excited. Returning to their ground state, electrons transition from higher to lower energy levels in the atom, emitting a photon of the same wavelength in a random direction. Since large amounts of photons are involved in the radiation processes with random directions, the total momentum remains zero, thus not affecting the motion of the radical as a whole. If we assume that this particular asymmetry of the CN coma is caused by an additional source, namely dust particles from the jets, then the deviation of the CN coma's axis of symmetry from the direction of the strongest jet becomes understandable. Additional acceleration by solar radiation should shift this axis in the antisolar direction, which is indeed observed.

The additional acceleration can be estimated by the fluorescence efficiency value, $L/N$, which varies with heliocentric velocity and heliocentric distance due to the Swings effect. Fluorescence efficiency $L/N$ for CN(0–0) at wavelength $\lambda 3883$ Å and the heliocentric distance of $r \approx 1.67$ au and heliocentric velocity (*r-dot*) of $v \approx 13$ km s$^{-1}$ of the comet is $L/N = 1.47 \times 10^{-13}$ erg s$^{-1}$ mol$^{-1}$ = $1.47 \times 10^{-20}$ J s$^{-1}$ mol$^{-1}$. $L/N = \Delta E/\Delta t$, where $\Delta E$ is the total photon energy that is re-emitted by one radical during the time $\Delta t$. For the time $\Delta t$, photons transmit momentum $\Delta P = \Delta E\, c^{-1}$. Considering that the velocity of light is $c \approx 3 \times 10^8$ m s$^{-1}$, and the mass of the radical C$^{12}$N$^{14}$ is $m = 4.32 \times 10^{-26}$ kg (https://www.webqc.org/molecular-weight-of-C12N14.html), then $\Delta P = L/N\, \Delta t\, c^{-1} = m\, \Delta v$, where $\Delta v$ is the change in the velocity of the molecule. Thus, the acceleration of the molecule is $a = \Delta v\, \Delta t^{-1} = L/N\, c^{-1}\, m^{-1} = 1.13 \times 10^{-3}$ m s$^{-2}$.

The value obtained can be considered a lower limit since only the CN (0–0) band is taken into account. To calculate the total acceleration, instead of the $L/N$ value for a single band, it is necessary to sum the $L/N$ values for all bands of the CN molecule, using the same formulas for calculating. If CN radicals originate from dust jets, then the acceleration due to the pressure of sunlight can blur and bend the shape of the CN jets. Thus, if different sections of the CN jet were formed with a time difference of 1 h, the additional shift between these sections would be approximately 1000 km.

CN radicals in comets originate mainly from i) the outgassing from the nucleus, ii) the dust particles in the near-nucleus coma, and iii) chemical processes in the coma. Since the discovery of the CN jets in comet Halley, A'Hearn et al. (1986) proposed that the CN seen in the jets could be produced by the dust particles escaping from the nucleus. The authors suggested that such particles could be submicrometer CHON grains discovered in comet 1P/Halley by space missions, from which CN molecules are directly produced. Combi and Delsemme (1980) showed that the CN radical can arise from the photolysis of HCN. The reliably detected parent molecule HCN in comet 1P/Halley (Despois et al., 1986; Schloerb et al., 1986; Winnberg et al., 1987) would seem to explain the origin of the CN radical. However, Bockelée-Morvan and Crovisier (1985) showed that the HCN production rate is less than the production rate of CN and, therefore, the CN radical cannot be associated only with the photodissociation of HCN. Several other molecules were proposed as parent molecules for CN radical, e.g., $C_2N_2$ (Bockelée-Morvan and Crovisier, 1985; Festou et al., 1998), $HC_3N$ or $C_4N_2$ (Krasnopolsky and Tkachuk, 1991). According to the ROSINA measurements (Hänni et al., 2020), significant quantities of CN radicals and their parent molecules HCN were constantly detected near the nucleus of comet 67P/C-G. There have also been cases where the quantity of CN radicals exceeded what could be produced from the observed amount of HCN molecules alone. Hänni et al. (2020) concluded that there is an additional source of CN originating from dust particles. The problem of the origin of cometary CN radicals has not yet been completely resolved and is still widely debated.

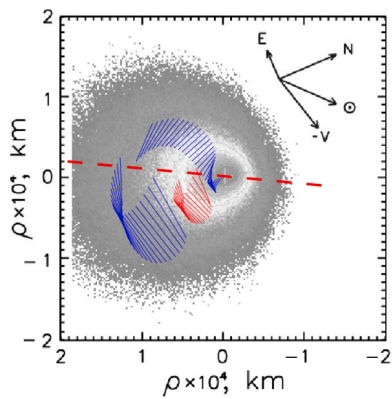

**Fig. 12.** Modeling the morphology of the CN coma in comet 67P/Churyumov–Gerasimenko observed on 6 February 2022, using a geometric model. The model jets are depicted as a set of blue and red lines, and the red dashed line shows the projection of the nucleus rotation axis onto the celestial plane. The negative distance is in the solar direction, and the positive distance is in the anti-solar direction. The cross and arrows are the same as shown and described in Fig. 3. (For interpretation of the references to color in this figure legend, the reader is referred to the web version of this article.)

## 5. Photometric characteristics of the comet

The broadband images in *g*-sdss and *r*-sdss filters were used to study the behavior of the production rate of the dust (in terms of *Afρ*) and its color and the normalized spectral gradient of the reflectivity as a function of the projected cometocentric distance ρ. We used imaging photometry to determine the apparent magnitude of the comet which is then used to compute the *Afρ*. Using calculated magnitudes of the coma, we estimated the color index *g–r*.

### 5.1. LT images of the comet

Fig. 13 (a, c) illustrates the average composite images of comet 67P/C−G taken in the *r*-sdss band with the 2-m Liverpool Telescope on 9 October 2021 and 6 February 2022. We applied the same techniques to selecting structures, which we used to process images obtained at the 6-m telescope. As can be seen in panels (b) and (d), the number, orientation, and shape of structures in the coma detected in the LT images of comet 67P/C−G taken in 2021 and 2022 are very similar to those revealed in images obtained with the 6-m telescope (Figs. 3 and 4). Some differences in the jets are most likely due to the rotation of the nucleus. The average *g–r* color of the cometary coma measured by the LT instrument across the observation period between 18 July 2015 and 11 June 2016 is $0.61^m \pm 0.004^m$ (Gardener et al., 2022). According to our calculation, the *g–r* color is $0.58^m \pm 0.04^m$ on 7 February 2022 (Table 5). These values are within the measurement errors and are consistent with corresponding estimates found at similar heliocentric distances (Jewitt and Meech, 1986; Boehnhardt et al., 2016).

**Table 5.**
Measured magnitudes, color, *Afρ* values, and normalized reflectivity gradient for comet 67P/Churyumov−Gerasimenko.

| Date UT | Filter | r [au] | m** [mag] | g–r [mag] | Afρ [cm] | A(0°)fρ*** [cm] | S' [% 1000 Å$^{-1}$] |
|---|---|---|---|---|---|---|---|
| 2021 Oct. 6 | g-sdss | 1.25 | 11.71 ± 0.01 | 0.63 ± 0.02 | 173 ± 1 | 806 ± 5 | 13.4 ± 0.3 |
| | r-sdss | 1.25 | 11.08 ± 0.01 | | 202 ± 4 | 943 ± 14 | |
| 2021 Oct. 9 | r-sdss | 1.25 | 11.29 ± 0.04 | | 159 ± 6 | 742 ± 28 | |
| 2022 Feb. 7 | g-sdss | 1.66 | 13.22 ± 0.02 | 0.58 ± 0.04 | 175 ± 6 | 255 ± 9 | 8.4 ± 1.9 |
| | r-sdss | 1.68 | 12.63 ± 0.02 | | 196 ± 5 | 286 ± 7 | |
| 2015 Nov. 8* | g-sdss | 1.61 | – | – | 184 ± 17 | 566 ± 52 | |
| 2015 Nov. 8* | r-sdss | 1.61 | – | – | 162 ± 18 | 498 ± 18 | |

\* results take from Rosenbush et al., 2017
\*\* are magnitudes of the comet with an aperture of radius ρ = 10,000 km.
\*\*\* The phase function is taken from Bertini et al., 2019

### 5.2. Dust production rate and normalized reflectivity gradient

The magnitude of the coma in the broadband filters was determined with a circular aperture centered on the central brightness peak found using the isophotes according to the following expression:

$$m_c(\lambda) = -2.5\lg\left[\frac{I_c(\lambda)}{I_s(\lambda)}\right] + m_s - 2.5\lg P(\lambda)\Delta M, \quad (2)$$

where $m_c$ and $m_s$ are the integrated magnitudes of the comet for the specific aperture of radius ρ and standard star, $I_c$ and $I_s$ are the measured fluxes of the comet and star, respectively; $P(\lambda)$ is the sky transparency that depends on the wavelength λ, and $\Delta M$ is the difference between the air masses of the comet and the star. The second term can be neglected since we used field stars as standards for broadband observations. The photometry results for an aperture of about 10,000 km are given in Table 5. An error in the magnitude measurements is $0.01^m$.

The activity level of comet 67P/C−G can be determined using the parameter *Afρ*, which is a measure of the dust production rate and is usually applied to characterize dust activity (A'Hearn et al., 1984). This parameter is the product of the albedo of cometary dust $A$, the filling factor $f$ defined as the mean cross-section of the particles per aperture surface area, and the projected aperture radius $\rho$. In the case of a stationary coma, i.e. an isotropic ejection of particles with a constant velocity and uniform expansion, this parameter does not depend on the aperture radius $\rho$. According to Mazzotta Epifani et al. (2011):

$$Af\rho = \frac{4r^2\Delta^2}{\rho} \times 10^{0.4(m_{Sun}-m_{comet})} \quad (3)$$

where $m_{comet}$ and $m_{Sun}$ are magnitudes of the comet and the Sun in the same filters, respectively, $r$ and $\Delta$ are heliocentric (in au) and geocentric (in cm) distances of the comet, $\rho$ is the aperture size (in cm). Magnitudes of the Sun for the corresponding filters were taken from Willmer (2018): $-26.47^m$ for the $g$-sdss and $-26.93^m$ for the $r$-sdss filters. For the calculation of $Af\rho$, the images obtained in the $g$-sdss and $r$-sdss filters on 6 October 2021, and the aperture of radius $\rho = 10{,}000$ km were taken. These values together with their errors are given in Table 5.

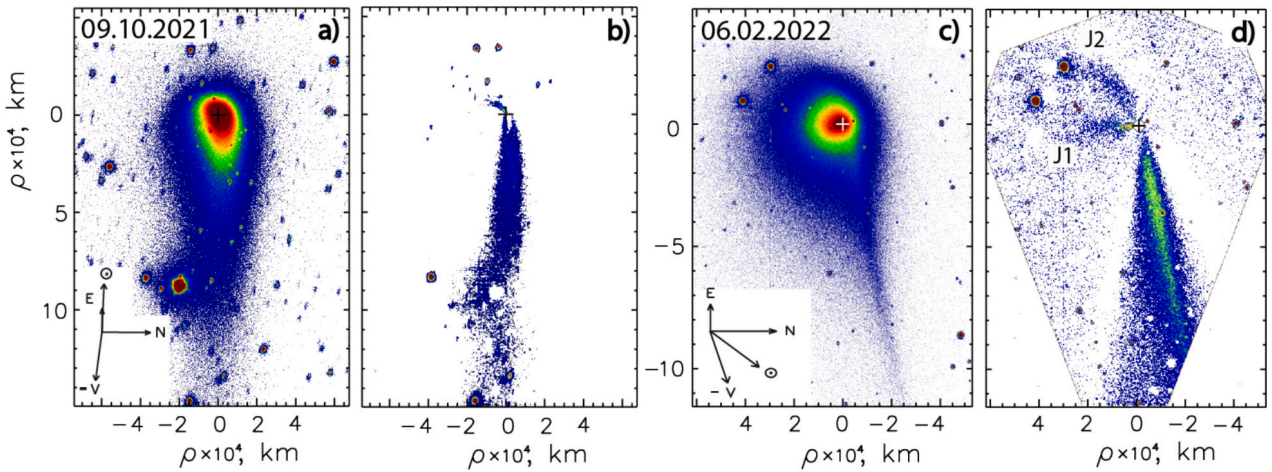

**Fig. 13.** Images of comet 67P/Churyumov−Gerasimenko taken with the 2-m Liverpool Telescope in the *r*-sdss filter on 9 October 2021 and 6 February 2022. Panels (a) and (c) are the direct CCD images of the comet with superimposed isophotes differing by a factor of $\sqrt{2}$ in intensity; (b) and (d) are the enhanced images processed by a rotational gradient technique of Larson and Sekanina (1984). Images are color-enhanced to show contrast. The cross and arrows are the same as shown and described in Fig. 3.

The variations of $Af\rho$ value with the cometocentric distance $\rho$ derived from observations of comet 67P/C−G on 6 October 2021 in the $g$-sdss and $r$-sdss filters were studied. We used apertures of increasing radius in 1-pixel increments, ranging from 0.39″ (which corresponds to 1 px or ∼ 134 km) to 74.3″ (∼ 25,561 km). Fig. 14 (left panel) presents the observed $Af\rho$ profiles of the comet in the $g$-sdss and $r$-sdss filters (without reduction to zero phase angle). The standard error in the measurements of $Af\rho$ is calculated using the errors of the parameters included in Eq. (3). According to the spectrum of comet 67P/C−G, the contribution from the molecular emissions to the $g$-sdss and $r$-sdss bands is 7.1 % and 3.6 %, respectively. Hence, the $Af\rho$ profiles should not change significantly due to the contribution of emissions, especially in the $r$-sdss band. $Af\rho$ values are higher in the $r$-sdss band than in the $g$-sdss band up to about 25,000 km. Excluding the seeing region, which is about 689 km, we can conclude that $Af\rho$ values first sharply increase up to the distances of about 2000 km, to ∼235 cm in the $g$-sdss filter and ∼ 290 cm in the $r$-sdss filter, and then gradually decrease to 25,000 km, ∼ 120 cm in both filters. The maximum value of $Af\rho \approx 290$ cm, obtained in an aperture with a radius of 2000 km (Fig. 14), exactly corresponds to the position of $Af\rho$ on the linear fitting curve at a heliocentric distance of $r = 1.25$ au in $\log Af\rho - \log r$ representation based on all available data for 1982, 1996, 2002, 2009, and 2015 apparitions of comet 67P/C−G (see Fig. 6 in Ivanova et al., 2017a). Fig. 14 (right panel) illustrates the variations of the *normalized spectral gradient* of the dust *reflectivity* $S'$ within the $g$-$r$ domain, calculated with different apertures, with a step size of 134 km, for observations of comet 67P/C−G on 6 October 2021. At the cometocentric distance of about 2000 km, the value of the normalized reflectivity gradient is about (13.4 ± 0.3) % per 1000 Å, which then decreases linearly with increasing distance from the nucleus. This value is close to the normalized reflectance gradient obtained from spectral observations, (13.6 ± 0.1) % per 1000 Å.

The similarity of the $Af\rho$ profiles in both filters suggests the dominance of dust in the comet, especially in the inner coma, which is confirmed by the spectrum of comet 67P/C−G (see Fig. 2). In the case of a steady state outflow of dust, the production rate of dust in the coma, in the sense of $Af\rho$, should be independent of the aperture size. However, since the $Af\rho$ in the inner coma, up to 2000 km, increases sharply, and then nonlinearly decreases with distance from the nucleus, it can be inferred that non-stationary processes take place in the comet (see for details Section 4. Morphology of dust and gas coma) or there is the dust with different physical properties in the inner and outer coma. The variations of the $Af\rho$ parameter with cometocentric distance in comet 67P/C−G resemble its behavior in both short-period comets (e.g. Tozzi et al., 2007; Mazzotta Epifani et al., 2011) and long-period comets (Mazzotta Epifani et al., 2014; Sun et al., 2024), as well as in distant comets (Korsun et al., 2010; Ivanova et al., 2019). The difference in the $Af\rho$ values in different filters is also similar to that we observed, for example, in comet C/2014 A4 (SONEAR) (Ivanova et al., 2019), which was explained by the wavelength-dependent variation in $Af\rho$ and the corresponding reddening trend (see Fig. 2b) which were attributed to changes in dust properties.

The variations of $Af\rho$ with cometocentric distance $\rho$ are accompanied by corresponding changes in the spectral gradient of dust reflectivity, which shows a sharp decrease $S'$ up to 2000 km, and then a gradual linear decrease with cometocentric distance. Both these effects may indicate two types of changes in the dust particles: fast change in the inner coma and smoother change in the outer coma. These changes are most likely due to the dynamical sorting effect, where the particle size distribution varies with cometocentric distance because smaller particles are more affected by radiation pressure. However, we do not exclude a possible contribution from particle fragmentation or compositional changes, such as the sublimation of volatiles.

### 5.3. Radial profiles of surface brightness

To investigate the overall distribution of cometary material as a function of distance from the nucleus, radial profiles were produced in each image by measuring brightness in a 3 × 3 pixel area along cuts through the coma. Having the intensity images of comet 67P/C−G in the $g$-sdss and $r$-sdss bands for 6 October 2021 and in the $BC$ and $RC$ narrowband comet filters for 6 February 2022, we constructed radial profiles of surface brightness using the technique described by Ivanova et al. (2019, 2021, 2023). Unfortunately, the BC and RC filters do not have photometric standards. Therefore, the transition to absolute fluxes was made using spectral observations performed at the same telescope. This means that the instrumental profile of the instrument for photometric and spectral observations is the same. As a result, the radial cross-cuts from the photometric center of the comet along the selected directions, namely, for jet structure, J1 and J2, dust tail, neckline, and coma, were obtained in all filters used for observations. For noise removal from images and to level out small-scale fluctuations in surface brightness, the profiles were smoothed with the 3 × 3 median filter. Strongly deviating values in individual pixels were replaced by the average of two adjacent ones. The uncertainty in the coma profiles was mainly due to the uncertainty in the brightness of the night sky which was determined by averaging the pixels in the region outside the comet not affected by the coma brightness. According to our estimates, the 1–σ brightness uncertainty is on average about 1 %.

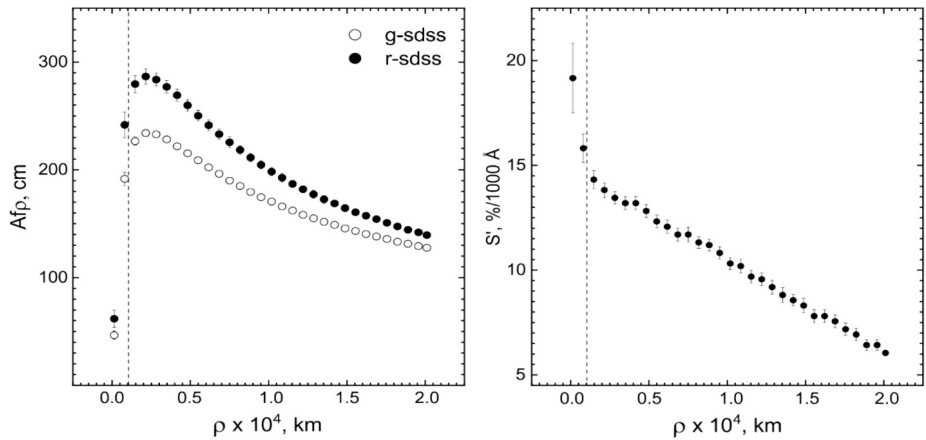

**Fig. 14.** $Af\rho$ profiles taken in the $g$-sdss and $r$-sdss filters as observed in comet 67P/Churyumov−Gerasimenko on 6 October 2021 (left panel). The right panel shows a change in the normalized spectral gradient of the reflectivity $S'$ within the $g$-$r$ domain with a cometocentric distance $\rho$ at the same date. The near-nucleus area, which may be affected by seeing, is delimited here by the vertical dashed line and was not considered.

Fig. 15 shows the radial surface brightness profiles of the comet 67P/C-G coma, determined from the images obtained at both the BTA and LT telescopes in the red filters, which best reflect the state of the dust coma because gas emissions weakly contaminate them. A detailed description of the profiles is given in the figure legend. On 6 October 2021, all profiles practically coincide up to cometocentric distances of about 15,000 km. Further, the slope of the tail profile does not change, while the brightness drops sharply in other profiles. On 6 February 2022, there is no significant difference between the slopes of the brightness profiles of both jets J1 and J2.

As it is known, the brightness changes with the distance from the nucleus according to the dependence $I \propto \rho^n$, where $I$ is the brightness of the coma at the cometocentric distance $\rho$ from the nucleus, and the exponent $n$ represents the dimensionless slope in the $\log I$–$\log \rho$ dependence. For a steady and isotropic ejection of long-living grains, the exponent should be $n = -1$. The slope of the linear fit to each profile, in log-log representation, is given in Table 6. The slopes of the profiles differ significantly for observations in October 2021 ($r \approx 1.25$ au before perihelion) and February 2022 ($r \approx 1.68$ au after perihelion). On average, the slope of the profiles before perihelion is approximately $-1.3$, while after perihelion it is about $-1$. As shown in Fig. 15 and Table 6, the brightness profiles are steeper before perihelion, most probably indicating enhanced dust production as the comet approaches perihelion.

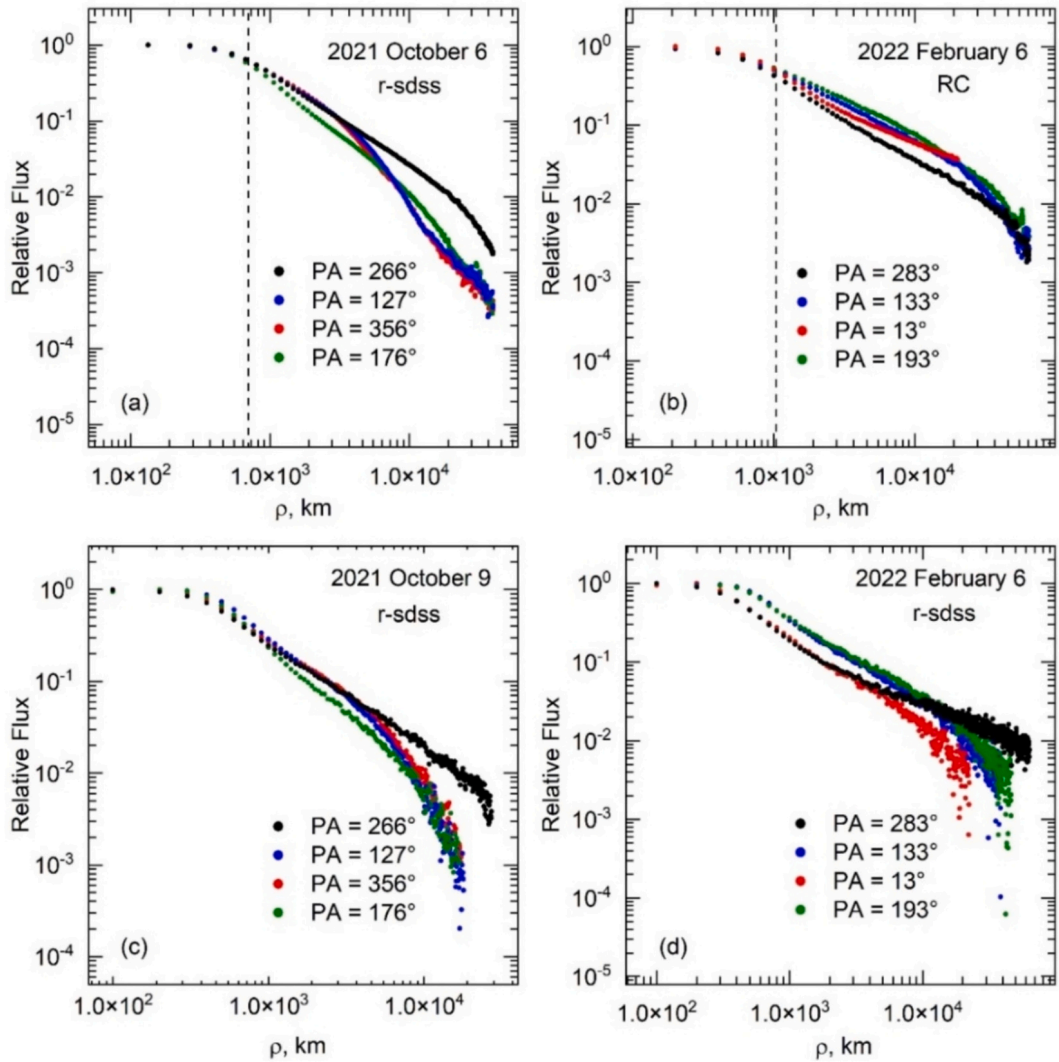

**Fig. 15.** Radial profiles of the surface brightness of the dust coma of comet 67P/Churyumov–Gerasimenko as a function of projected cometocentric distance $\rho$ obtained from the images taken in the *r-sdss* filters on 6 October 2021 (left panel) and in the *RC* continuum filters observed on (right panel) at the 6-m telescope (top row). The bottom row presents the same profiles obtained from the images taken in the *r-sdss* filters on 9 October 2021 (left panel) and 6 February 2022 (right panel) at the Liverpool Telescope. The individual curves are cross-cuts measured from the photometric center of the coma along specific directions and designated as follows: in the left panels, the black line is the cross-cut along the tail with $PA = 266°$, red and green lines are the scan across the coma perpendicular to the solar-antisolar direction in the directions $PA = 356°$ and $PA = 176°$, respectively, and blue line is the scan along the jet in the direction $PA = 127°$. In the right panels, the black line is directed along the neckline in the solar direction ($PA = 283°$); the red line passes through the coma in the direction $PA = 13°$ opposite to the direction of the jet J2 with $PA = 193°$ which is shown by the green line. The blue line points out the direction of the jet J1 in the anti-solar direction, $PA = 133°$. The near-nucleus area, which may be affected by seeing, is delimited here by the vertical dashed line and was not considered. (For interpretation of the references to color in this figure legend, the reader is referred to the web version of this article.)

## 6. Comparison of comet 67P/C-G in the 2015/16 and 2021/22 apparitions

### 6.1. Spectra

In the 2015/16 apparition, spectra of comet 67P/C-G were obtained at the 6-m BTA telescope with multi-mode focal reducer SCORPIO-2 after its perihelion on 13 August 2015, between November 2015 and April 2016 at a heliocentric distance $r$ from 1.6 to 2.7 au (Ivanova et al., 2017a, 2017b). Strong CN and relatively weak $C_2$, $C_3$, and $CO^+$ emissions were identified in the spectrum of the comet on 8 November and 9 December 2015. The numerous weak emissions of $NH_2$ were only observed in November. Only CN emission was detected on 3 April 2016. This means that comet activity decreased as it moved away from the Sun. Therefore, the production rate $Q$ of CN molecules also decreased from $7.05 \times 10^{24}$ mol s$^{-1}$ at $r = 1.62$ au in November 2015 to $5.3 \times 10^{23}$ mol s$^{-1}$ at $r = 2.72$ au in April 2016. Spectral observations in the 2021/22 apparition were performed at the same telescope in the post-perihelion epoch at $r = 1.68$ au. The production rate $Q(CN) = 1.7 \times 10^{24}$ mol s$^{-1}$ at this heliocentric distance is approximately 4 times less compared to the value of $Q(CN)$ in the 2015/16 apparition. Overall, comet 67P/C-G showed a lower value of the gas production rate in 2021/22 than in 2015/16.

This result appears to be inconsistent with the main conclusions on the activity evolution reported by Bair et al. (2022), who found higher production rates in 2022. However, it is difficult to draw firm conclusions from a single point of comparison, given the slight differences in heliocentric distance and the timing of perihelion passage observations. It should also be taken into account that, against the background of the rapid decline in cometary activity at this orbital phase and the diurnal modulation caused by the nucleus rotation, variations in the comet's production rates may occur.

**Table 6.**
Radial slopes of profiles $n$, in a log–log scale, derived from the measured radial profiles of the surface brightness of comet 67P/Churyumov–Gerasimenko along selected directions: tail, neckline, jets, and coma.

| Structure | Position angle [deg] | $\rho$ [km] | $n$ | |
|---|---|---|---|---|
| | | | BTA | LT |
| 6–9 October 2021 | | | | |
| Tail | 266 | 700–11,000 | $-1.23 \pm 0.03$ | $-1.16 \pm 0.01$ |
| Jet2 | 133 | 700–3500 | $-1.23 \pm 0.03$ | $-1.36 \pm 0.01$ |
| Coma | 356 | 700–5000 | $-1.17 \pm 0.02$ | $-1.18 \pm 0.01$ |
| Coma | 176 | 700–8000 | $-1.45 \pm 0.01$ | $-1.53 \pm 0.02$ |
| | | | | |
| 6 February 2022 | | | | |
| Neckline | 283 | 1000–15,000 | $-0.94 \pm 0.01$ | $-1.03 \pm 0.01$ |
| Jet1 | 133 | 1000–11,000 | $-0.94 \pm 0.01$ | $-1.03 \pm 0.01$ |
| Jet2 | 193 | 1000–11,000 | $-0.96 \pm 0.01$ | $-1.00 \pm 0.01$ |
| Coma | 13 | 1000–11,000 | $-0.92 \pm 0.01$ | $-1.10 \pm 0.01$ |

The obtained $\log[Q(C_2/CN)]$ ratios for both apparitions of comet 67P/C-G, at approximately the same post-perihelion distance from the Sun $r = 1.6$ au, are typical for carbon-chain depleted comets. Indeed, these ratios are $-0.43$ on 8 November 2015 and $-0.60$ on 6 February 2022, characteristic of comets with low carbon abundance. However, it should be noted that Le Roy et al. (2015) showed that there is a large difference in the relative abundance of carbon-chain molecules between the summer and winter hemispheres related to different seasonal illumination of the nucleus. Therefore, it is impossible to draw a reasonable correct conclusion about whether this comet is normal or carbon-chain depleted. According to A'Hearn et al. (1995), Schleicher (2008), and Fink (2009), carbon-chain-depleted comets may be older and more evolved objects. Processes such as degassing, thermal effects, and solar radiation can reduce the content of volatile carbon compounds such as $C_2$ and $C_3$. Comet 67P/C-G, classified as a carbon-chain depleted comet, likely underwent similar processes during its evolution.

The spectra of comet 67P/C-G in both apparitions show a high continuum consistent with dust-rich comets and an increase in dust reflectivity with increasing wavelength. The normalized reflectivity gradient $S'$ (or normalized reddening) of the dust was on average about 16 % 1000 Å$^{-1}$ on 6 October 2021 and about 14 % 1000 Å$^{-1}$ on 6 February 2022 within the wavelength interval of $\lambda 4450$–6839 Å and at heliocentric distances 1.25 au and 1.68 au, respectively. In November 2015 and April 2016, $S'$ slightly varied from 12.5 to 11.3 % 1000 Å$^{-1}$ in the range $\lambda 3600$–7070 Å at $r = 1.62$ and 2.72 au, respectively. The obtained values of spectral gradient indicate that the dust particles in the comet were larger than the wavelength. The slight difference in dust reddening in comet 67P/C – G may be due to the difference in phase angles of observations.

Dust production. The 2015/2016 period was characterized by moderate dust activity of comet 67P/C-G. $Af\rho$ values normalized to zero phase angle show a decrease in activity with increasing heliocentric distance: in the r-sdss filter, $Af\rho$ decreased from approximately $498 \pm 18$ cm at 1.61 au to significantly lower values (72 cm $\pm$ 7 cm) at 3.9 au (Rosenbush et al., 2017). Observations in 2021/2022, also normalized to $A(0°)f\rho$, revealed higher dust activity: for example, in the g-sdss filter, $Af\rho$ reached $806 \pm 5$ cm on 6 October 2021, significantly exceeding the 2015 value ($566 \pm 52$ cm). However, by February 2022, the activity had decreased substantially to $255 \pm 9$ cm. A similar trend was observed in the r-sdss filter: $Af\rho$ decreased from $943 \pm 14$ cm in October 2021 to $286 \pm 7$ cm in February 2022, whereas it was $498 \pm 18$ cm in 2015.

### 6.2. Radial brightness profiles

Comparison of the surface brightness profiles of comet 67P/C-G in previous (Rosenbush et al., 2017) and subsequent apparitions highlights the complex and heterogeneous nature of the coma, requiring more observations over a longer period for modeling and detailed analysis of the profiles. The slopes of the brightness profiles vary depending on the direction, namely tail, jets, or coma, assuming a non-isotropic ejection of dust particles from the whole surface of the nucleus as well as from local areas on the nucleus. In this study, we compared the profiles before (October 2021) and after (February 2022) perihelion, showing how the profile slopes change over time. For example, radial slopes are significantly larger than unity before perihelion, whereas after perihelion the slopes become closer to $-1$, indicating more stable dust emission. Summarizing, it should be noted that the radial brightness profiles in 2015–2016 were flatter at small distances and steeper with increasing distance from the nucleus. In contrast, the profiles in 2021–2022 demonstrate a more gradual alteration in the shape of the curve with cometocentric distance and less pronounced fluctuations. For both periods of observations of the comet, it is evident that the tail profile significantly differs from that of the coma and other structures, yet remains relatively stable across both apparitions of the comet.

*6.3. Morphology of the dust coma*

To compare the jet activity in both apparitions of comet 67P/C-G, we analyzed images taken with the 6-m BTA telescope on 8 November 2015 (Rosenbush et al., 2017). The raw image of the comet in the r-sdss broadband filter is shown in Fig. 16a, and (b) shows the image of the comet after applying a rotational gradient filter with superimposed model jets. Using a geometric model, we determined the latitudes of the active areas that form the J1 and J2 jets (panel b). The first active area producing the J1 jet is located at a latitude of $\varphi_1 = -85° \pm 5°$, and the second one at $\varphi_2 = -30° \pm 8°$. Panel (c) demonstrates the possible location of active areas on the cometary nucleus. Notably, the Rosetta mission's study of comet 67P/Churyumov–Gerasimenko revealed that latitude $\varphi_1$ corresponds to the highly active Bes region (Fornasier et al., 2017). Latitude $\varphi_2$ is close to the latitude of the subsolar point on that date ($\varphi_{sun} = -24°$). According to Tubiana et al. (2019), the maximum emission of gas and dust consistently aligns approximately with the direction of the Sun.

Comparing the parameters of the jets in different observation epochs, it is clear that the coordinates of the active areas observed in 2015 do not coincide with either our data or with the data obtained by Boehnhardt et al. (2024) from observations of comet 67P/C-G in 2021–2022. This suggests that there are no long-term stable jet sources at the nucleus of comet 67P/C-G, although this contradicts the conclusion reached by Boehnhardt et al. (2024), who observed the same patterns in 2002, 2009, 2015, and 2021 apparitions. Similar behavior, with shifting of the jet source regions, was noted during studies of the nucleus by the *Rosetta* mission. However, some comets exhibit more stable activity. For example, comet 2P/Encke has maintained two active jet-producing regions for several decades (Sekanina, 1988; Rosenbush et al., 2020).

## 7. Conclusions

In the first part of our study, we analyzed the results of the photometric and spectroscopic observations of comet 67P/C-G obtained during its latest apparition in 2021/22. We compared them with the corresponding results from the 2015/16 observations. In both apparitions, quasi-simultaneous photometric and spectroscopic observations of the comet were carried out at the 6-m BTA SAO telescope. For comparison, we supplemented these data with images of the comet taken with the 2-m Liverpool Telescope.

Throughout this paper, we drew several conclusions regarding the properties of the observed coma of comet 67P/C-G:

1. The spectra of comet 67P/C-G show a high continuum level, consistent with dust-rich comets, and a fairly strong CN and relatively weak $C_3$, $C_2$, and $NH_2$ emissions. Overall, the comet showed a lower value of the gas production rate in the 2021/22 apparition than in 2015/16. Comet 67P/C-G has been confirmed to be classified as a carbon-depleted comet, with a decrease in volatile carbon content observed over time.

2. The comet's coma in 2021 exhibited a higher dust production level than in 2015/2016 when $Af\rho$ values were corrected and normalized to zero phase angle. In the g-sdss filter, $Af\rho$ was approximately $806 \pm 5$ cm on 6 October 2021, compared to $566 \pm 52$ cm on 8 November 2015. However, during the 2021/2022 apparition, the dust activity decreased significantly to about $255 \pm 9$ cm in the g-sdss filter on 7 February 2022. Similar trends are observed in the r-sdss filter, with $Af\rho$ dropping from $943 \pm 14$ cm (6 October 2021) to $286 \pm 7$ cm (7 February 2022), while the 2015 value was $498 \pm 18$ cm. Radial $Af\rho$ profiles indicate a decline in dust production with increasing distance from the nucleus, which may suggest dynamical sorting of dust particles. The normalized reflectivity gradient of the dust averaged approximately 16 % per 1000 Å on 6 October 2021 and about 14 % per 1000 Å on 6 February 2022 within the wavelength range of 4450–6839 Å.

3. The average g–r color of the cometary coma is $0.63^m \pm 0.02^m$ on 6 October 2021 and $g–r = 0.58^m \pm 0.04^m$ on 6 February 2022. The measured g–r color with the Liverpool Telescope across the observation period between 18 July 2015 and 11 June 2016 is $0.61^m \pm 0.004^m$ (Gardener et al., 2022). All these values are within the measurement errors and consistent with corresponding estimates at similar heliocentric distances.

4. For both periods of observations of the comet, the surface brightness profiles of the tail significantly differ from those of the coma and jets, yet remain relatively stable across both apparitions. The slopes of the profiles differ significantly for observations in October 2021 ($r = 1.25$ au before perihelion) and February 2022 ($r = 1.68$ au after perihelion). On average, the slope of the profiles before perihelion is approximately $-1.3$, while after perihelion it is about $-1$.

5. The enhanced images of comet 67P/C−G obtained pre-perihelion on 6 October 2021 show in both the g-sdss and r-sdss filters a jet-like structure directed toward the Sun at the position angle $PA_{jet} \approx 127°$ and a dust tail in the antisolar direction at $PA_{tail} \approx 266°$. Using a geometric model to simulate the jet formation process, the latitude of the active area on the nucleus $\varphi = -70° \pm 4°$, the jet opening angle of $20° \pm 6°$, and the particle ejection velocity of $0.30 \pm 0.04$ km s$^{-1}$ are determined.

6. After perihelion, on 6 February 2022, the structure of the dust coma changed. Morphology analysis clearly shows the neckline structure in the solar direction ($PA_{NL} = 284°$), two jet-like structures, one of which is approximately perpendicular to the solar-antisolar direction at $PA_{J1} = 193°$, and the second is directed approximately in the antisolar direction at $PA_{J2} = 133°$. A geometric model predicts that both jets J1 and J2 originate from the same active source on the rotating nucleus located at a latitude of $\varphi = -58° \pm 5°$ with the opening angle of the jet $26° \pm 8°$. Moreover, we have shown that there is another active source located at latitude $\varphi = -53° \pm 10°$ and is separated from the first one by a difference in longitude of $150° \pm 20°$. The second active area generates the J3 jet which overlaps jet J1 and fills the gap that is created by the activity of the first active area. The average particle velocity in the jets is estimated to be about $0.32 \pm 0.04$ km s$^{-1}$. The suspected first active region of the outflow of the jets predominantly lies on a convex part of the nucleus surface near the border of formations Khonsu and Bes, while the second region may be located in a depression near the border of Anhur and Bes, receiving less sunlight than the first region.

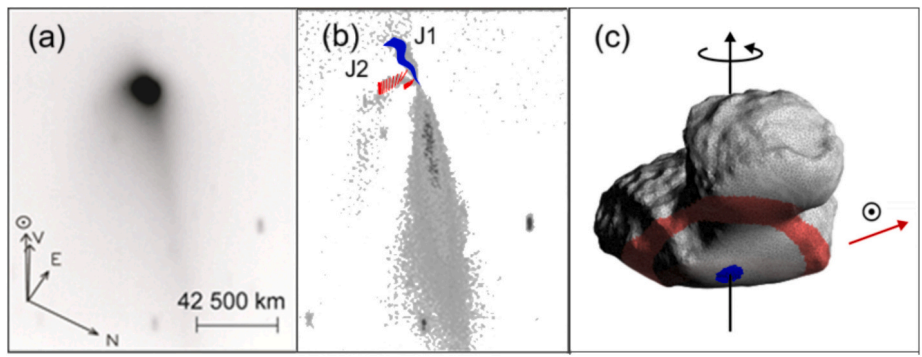

**Fig. 16.** Image of comet 67P/Churyumov−Gerasimenko taken with the 6-m BTA telescope through the *r*-sdss filter on 8 November 2015. Panel (a) is the raw image of the comet, and (b) is an image to which the rotational gradient technique of Larson and Sekanina (1984) is applied. This image is superimposed with jets modeled using a geometric model. (c) demonstrates the appearance of the comet nucleus on 8 November 2015, indicating the region from which the jet may originate. The colors of these areas correspond to those of the modeled jets in (b). The rest of the designations are the same as in Fig. 7.

7. Exploration of the sunward structure using the geometric and Finson-Probstein models has shown that this structure is a true neckline (frequently observed in the coma of comet 67P/C−G in other apparitions), which is superimposed on the projection (not the true dust tail) of the dust tail onto the sky and is therefore observed in the solar direction. The observed neckline structure is primarily formed due to the ejection of large dust particles near the comet's perihelion on 2 November 2021, that is, more than 3 months before our observations on 6 February 2022.

8. The CN coma observed in images of comet 67P/C−G on 6 February 2022 through the cometary CN ($\lambda$3870/58 Å) filter looks like an asymmetric coma with a bright compact spot around the nucleus, up to ~5000 km, which may indicate a rapid release of a CN parent material. The CN coma is roughly elongated in the direction between the two dust jets J1 and J2, however, there is a noticeable shift from the direction opposite to the Sun, which suggests an additional source of CN radicals in the coma. An additional source of CN may be dust jets.

**CRediT authorship contribution statement**

**Vira Rosenbush:** Writing – review & editing, Writing – original draft, Supervision, Project administration, Methodology, Formal analysis, Conceptualization. **Valerii Kleshchonok:** Writing – review & editing, Visualization, Software, Investigation, Formal analysis. **Oleksandra Ivanova:** Writing – review & editing, Project administration, Methodology, Formal analysis, Data curation, Conceptualization. **Igor Luk'yanyk:** Writing – review & editing, Visualization, Software, Investigation, Formal analysis. **Colin Snodgrass:** Writing – review & editing, Supervision, Methodology, Data curation, Conceptualization. **Daniel Gardener:** Data curation. **Ludmilla Kolokolova:** Formal analysis, Conceptualization. **Johannes Markkanen:** Writing – review & editing, Methodology, Formal analysis, Conceptualization. **Elena Shablovinskaya:** Data curation.

**Declaration of competing interest**

The authors declare that they have no known competing financial interests or personal relationships that could have appeared to influence the work reported in this paper.

**Acknowledgments**

This work is supported by the Partnership Fund of the University of Edinburgh and Taras Shevchenko National University of Kyiv. The work by OI is supported by the Slovak Grant Agency for Science VEGA (Grant No. 2/0059/22). Research of VR and IL is supported by Project No. 0125U002319 of the Ministry of Education and Science of Ukraine. Also, the research of IL and VK was supported by the project of the Ministry of Education and Science of Ukraine No. 0124U001304. JM was supported by the German Research Foundation (DFG) grant no. 517146316.

**Appendix A. Aligning images with sub-pixel accuracy**

CCD observations of comets typically produce a set of images which must then be stacked together. The main problem with stacking frames is that it is almost impossible to obtain the image of the comet in the same position within all frames due to the motion of the comet, inaccurate telescope tracking at long exposures, refraction, and other reasons. This effect can have different magnitudes.

It is relatively easy to account for integer-pixel shifts in the position of a comet. However, if the image needs to be shifted by a fraction of a pixel, then a special technique must be applied. Aligning a set of frames with comets to a standard position with sub-pixel accuracy is important for various applications, such as increasing the signal-to-noise ratio to explore the physical properties of comets, subtracting frames to obtain a color map, creating a polarization map, studying temporary changes in the morphological structures of the coma, etc.

First, the pixel with the highest brightness is determined in the coma, which is considered to be an approximation of the optocenter. After this, two spatial profiles are generated, one of them along the X-axis and the other along the Y-axis, passing through the pixel with the maximum brightness. Using these profiles, we determine the precise position of the optocenter with precision down to a fraction of a pixel. For a profile along the X-axis, we assumed that the surface brightness of the coma is determined according to the following expression:

$$i(x,y) = \begin{cases} \dfrac{I_0}{(x^2+y^2)^{n^-} + a}, & x < 0 \\ \dfrac{I_0}{(x^2+y^2)^{n^+} + a}, & x \geq 0 \end{cases}, \quad i(x,y) = \begin{cases} \dfrac{I_0}{(x^2+y^2)^{n^-} + a}, & x < 0 \\ \dfrac{I_0}{(x^2+y^2)^{n^+} + a}, & x \geq 0 \end{cases}, \tag{A.1}$$

where $(x, y)$ are coordinates in pixels which are measured from the position of the cometary nucleus, $a$ is a constant which is approximately equal to the seeing value, $n^-$ is the exponent for the part of the spatial profile with negative values of the $x$ coordinate, $n^+$ is the exponent for the part of the profile with positive $x$ values. The different values of the exponents for the two parts of the profile take into account the asymmetry in the spatial distribution of brightness which is often observed in comas. The intensities of individual pixels are obtained by integrating the model brightness distribution over the pixel area:

$$I(x_k) = \int_{-d/2}^{d/2} \int_{x_k-d/2}^{x_k+d/2} i(x+\delta_x, y)dxdy, \tag{A.2}$$

where $I(x_k)$ is the intensity of the $k$-th pixel in the spatial profile along the X-axis, and $\delta_x$ is the shift of the optocenter along the X-axis from the center of the pixel with the maximum brightness. The parameters of the model profile, namely $a$, $n^-$, $n^+$, and $\delta_x$, are determined from the spatial brightness profile by the least squares method. An example of such determination of parameters is illustrated in Fig. A.1.

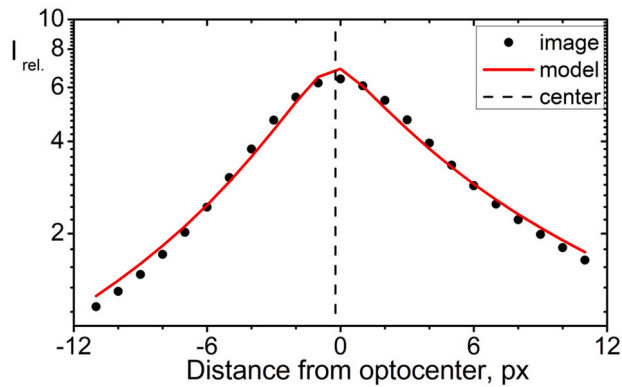

**Fig. A.1.** Determination of the exact position of the optocenter from the spatial brightness profile. The circles denote the measured pixel *intensity,* and the solid line shows the model profile.

The same procedure is applied to the spatial brightness profile along the Y-axis. As a result, we obtain the value of the optocenter displacement from the pixel center, denoted as ($\delta x$, $\delta y$). This shift corresponds to the offset of the entire image relative to the pixel grid. Using linear two-dimensional interpolation, we find the shift value for an image aligned to the pixel grid. All frames undergo this process. Thus, we obtain a set of frames aligned to the same position with sub-pixel accuracy.

## Data availability

Data will be made available on request.